\documentclass[sigconf]{acmart}

\AtBeginDocument{%
  }

\copyrightyear{2025}
\acmYear{2025}
\setcopyright{acmlicensed}
\acmConference[CHI '25]{CHI Conference on Human Factors in Computing Systems}{April 26-May 1, 2025}{Yokohama, Japan}
\acmBooktitle{CHI Conference on Human Factors in Computing Systems (CHI '25), April 26-May 1, 2025, Yokohama, Japan}
\acmDOI{10.1145/3706598.3713801}
\acmISBN{979-8-4007-1394-1/25/04}

\acmSubmissionID{7990}

\usepackage{enumitem}
\usepackage{multirow}
\usepackage{xspace}
\usepackage{url}
\newcommand{\tool}{\emph{SketchFlex}\xspace}

\usepackage[nonumberlist,nopostdot]{glossaries}
\setglossarystyle{altlist}
\makenoidxglossaries

\newglossarystyle{twocolborderbold}{
    {
    \begin{longtable}{p{0.3\textwidth}p{0.65\textwidth}}
    \noalign{\hrule height 1.5pt} % This will add a horizontal line at the top of the table
    }
    {
    \end{longtable}
    }%
  \renewcommand*{\subglossentry}[3]{%
  }%
}

\begin{document}

\newcommand{\eg}{\emph{e.g.}}
\newcommand{\ie}{\emph{i.e.}}

\newcommand{\strike}[1]{\textcolor{red}{\sout{#1}}}
\newcommand{\strikeg}[1]{\textcolor{blue}{\sout{#1}}}
\newcommand{\add}[1]{\textcolor{red}{#1}}
\newcommand{\replace}[2]{\strikeg{#1 }\add{#2}}
\newcommand{\new}[1]{\textcolor{blue}{#1}}
\newcommand{\cmt}[2]{\textcolor{blue}{#1}\add{#2}}
\newcommand{\yl}[1]{\textcolor{orange}{Ye: #1}}

%For the table of prompt
\definecolor{newgreen}{rgb}{0.0, 0.5, 0.0}
\definecolor{newblue}{rgb}{0.0, 0.0, 0.0}
\definecolor{newred}{rgb}{1.0, 0.0, 0.0}

\definecolor{blue}{rgb}{0.0, 0.0, 1.0}
\newcommand{\rev}[1]{\textcolor{blue}{#1}}

\newenvironment{tightcenter}{%
  \setlength\topsep{0pt}
  \setlength\parskip{0pt}
  \begin{center}
}{%
  \end{center}
}

\title[SketchFlex]{SketchFlex: Facilitating Spatial-Semantic Coherence in Text-to-Image Generation with Region-Based Sketches}

%%
%% The "author" command and its associated commands are used to define
%% the authors and their affiliations.
%% Of note is the shared affiliation of the first two authors, and the
%% "authornote" and "authornotemark" commands
%% used to denote shared contribution to the research.

\author{Haichuan Lin}
\affiliation{%
  \department{Thrust of Computational Media and Arts}
  \institution{The Hong Kong University of Science and Technology (Guangzhou)}
  \city{Guangzhou}
  \country{China}
}
\email{hlin386@connect.hkust-gz.edu.cn}

\author{Yilin Ye}
\authornote{Yilin Ye is the corresponding author.}
\affiliation{%
  \department{Thrust of Computational Media and Arts}
  \institution{The Hong Kong University of Science and Technology (Guangzhou)}
  \city{Guangzhou}
  \state{Guangdong}
  \country{China}
}
\affiliation{%
  \department{Academy of Interdisciplinary Studies}
  \institution{The Hong Kong University of Science and Technology}
  \city{Hong Kong SAR}
  \country{China}
}
\email{yyebd@connect.ust.hk}

\author{Jiazhi Xia}
\affiliation{%
  \department{School of Computer Science and Engineering}
  \institution{Central South University}
  \city{Changsha}
  \country{China}
}
\email{xiajiazhi@csu.edu.cn}

\author{Wei Zeng}
\affiliation{%
  \institution{The Hong Kong University of Science and Technology (Guangzhou)}
  \city{Guangzhou}
  \state{Guangdong}
  \country{China}
}
\affiliation{%
  \institution{The Hong Kong University of Science and Technology}
  \city{Hong Kong SAR}
  \country{China}
}
\email{weizeng@hkust-gz.edu.cn}

%%
%% By default, the full list of authors will be used in the page
%% headers. Often, this list is too long, and will overlap
%% other information printed in the page headers. This command allows
%% the author to define a more concise list
%% of authors' names for this purpose.
% \renewcommand{\shortauthors}{}
\renewcommand{\shortauthors}{Lin et al.}

\begin{abstract}
Text-to-image models can generate visually appealing images from text descriptions. 
Efforts have been devoted to improving model controls with prompt tuning and spatial conditioning. 
However, our formative study highlights the challenges for non-expert users in crafting appropriate prompts and specifying fine-grained spatial conditions (e.g., depth or canny references) to generate semantically cohesive images, especially when multiple objects are involved. 
In response, we introduce \tool, an interactive system designed to improve the flexibility of spatially conditioned image generation using rough region sketches. 
The system automatically infers user prompts with rational descriptions within a semantic space enriched by crowd-sourced object attributes and relationships. 
Additionally, \tool refines users' rough sketches into canny-based shape anchors, ensuring the generation quality and alignment of user intentions. 
Experimental results demonstrate that \tool achieves more cohesive image generations than end-to-end models, meanwhile significantly reducing cognitive load and better matching user intentions compared to region-based generation baseline.

\end{abstract}

%%
%% The code below is generated by the tool at http://dl.acm.org/ccs.cfm.
%% Please copy and paste the code instead of the example below.
%%


%%
%% Keywords. The author(s) should pick words that accurately describe
%% the work being presented. Separate the keywords with commas.

% \received{20 February 2023}
% \received[revised]{12 March 2009}
% \received[accepted]{5 June 2009}

%%
%% This command processes the author and affiliation and title
%% information and builds the first part of the formatted document.

\maketitle
\section{Introduction}\label{sec:introduction}
% text2image shows success in different areas
The rapid development of text-to-image (T2I) generative models, such as Midjourney~\cite{MID}, Stable Diffusion~\cite{rombach2022high}, and DALL-E~\cite{ramesh2022hierarchical}, has demonstrated impressive capabilities in generating high-quality, visually appealing images from text input.
These T2I models have been successfully utilized in various fields, including industrial design~\cite{liu20233dall}, visualization design~\cite{xiao2023let}, and visual art creation~\cite{ko2023large}.

% challenge for region control
\begin{figure*}[t]
    \centering
    \includegraphics[width=0.995\textwidth]{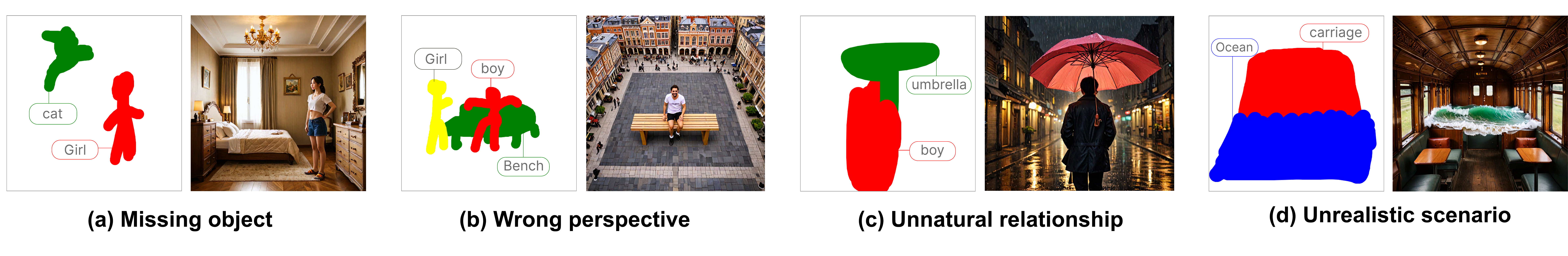}
    \vspace{-1mm}
    \caption{Failure cases of existing methods for rough sketch based image generation: a) missing object for the green sketch, b) wrong perspective of the man sitting on the bench, c) unnatural relationship as the man is not holding the umbrella, and d) unrealistic scenario for water in the carriage. (a)-(c) are generated by Dense Diffusion~\cite{kim2023dense}, (d) is generated by MultiDiffusion~\cite{bar2023multidiffusion}.}
    \Description{Failure cases of existing methods for rough sketch based image generation: a) missing object for the green sketch, b) wrong perspective of the man sitting on the bench, c) unnatural relationship as the man is not holding the umbrella, and d) unrealistic scenario for water in the carriage. (a)-(c) are generated by Dense Diffusion~\cite{kim2023dense}, while (d) is generated by MultiDiffusion~\cite{bar2023multidiffusion}.}
    \label{fig:problem_identify}
    \vspace{0mm}
\end{figure*}

To better control the outcomes of the T2I models, prompt-tuning methods have been proposed to help craft more effective prompts to generate the desired images~\cite{guo2024prompthis,wang2024promptcharm,feng2023promptmagician}. 
However, these models still confine users to abstract and sometimes ambiguous text modality interaction, failing to provide detailed spatial control of image composition~\cite{zhang2023adding}.
In response, researchers have developed various spatial conditioning techniques to allow for more refined control beyond prompt-based methods, including line sketches and color blocks, as well as depth maps, segmentation maps, and pose skeletons commonly used in computer vision~\cite{zhang2023adding,mou2024t2i,koley2023picture, kodaira2023streamdiffusion,lee2024streammultidiffusion}. 
These fine-grained spatial controls enable users to more easily translate their creative intent into images, contributing to the widespread adoption of these tools within the generative AI (GenAI) community.

% fine-grained control benefits, but not friendly to novice
There are differences in the way various user groups use AI tools~\cite{shi2023understanding}.
Although experienced users find the existing fine-grained spatial control models highly beneficial, these models present challenges for novice users.
In particular, novice users often encounter difficulties in attempting to create fine-grained spatial controls such as line sketches, depth maps, and semantic segmentation maps.
This challenge hinders novice users from effectively utilizing advanced control techniques to express their creative intentions.
In contrast, the rough sketch control~\cite{kim2023dense,bar2023multidiffusion,wang2024instancediffusion} offers a more accessible way for novice users to interact with. 
These methods allows users to roughly sketch regions within an image and assign specific prompts to those regions, ensuring the generated image aligns with the user's sketch. 
Existing works, such as StreamMultiDiffusion~\cite{lee2024streammultidiffusion}, incorporate rough sketch control with latent consistency models~\cite{luo2023lcm} to enable real-time painting effects, making it easier for novice users to express their creative intent through spatial control.

However, while rough sketch control offers convenience, it also introduces several challenges.
First, rough sketch control still requires the appropriate use of text prompts to guide generative models in producing desired images.
Constructing prompts that closely align with the rough sketch presents a significant challenge (\textbf{C1}).
In particular, novice users often struggle to articulate the relationships between objects.
Second, rough sketches created by novice users are typically of low quality.
Relying solely on these rough sketches can result in semantically incoherent images, particularly for situations where multiple objects are involved~\cite{bar2023multidiffusion}.
Potential issues include missing objects, wrong perspective, unnatural object relationships, and unrealistic scenarios, as shown in Figure~\ref{fig:problem_identify}.
The process of refining rough sketches into high-quality spatial conditions is nontrivial (\textbf{C2}).
Moreover, current models operate primarily in an end-to-end generation approach, which can be challenging to generate fully satisfactory results in a single iteration and does not align with the iterative design process~\cite{li2024realtimegen,huang2024plantography}. 
It is not uncommon for one aspect of the image to meet the user's expectations while another falls short, making iterative refinement highly desirable.
However, the lack of an appropriately designed interactive process can hinder iterative refinement (\textbf{C3}).

% facing these chanllenge, for novice user to create their work with flexible control condition, we propose...
To address these challenges, we develop \tool, an interactive system designed to empower novice users to flexibly create semantically cohesive images from rough sketches and prompts, with key modules as follows: 
\begin{enumerate}
\item \emph{Sketch-aware prompt recommendation} (Sect.~\ref{ssec:prompt_rec}) for refining prompts aligned with the rough sketch (\textbf{C1}).
Specifically, the system employs a multimodal large language model (MLLM) to interpret user prompts and sketches, generating content within a semantic space designed for multi-region spatial conditioning to ensure completeness across all regions. The prompt generation is grounded in semantic knowledge drawn from fine-grained visual attributes, states, and relationships of common objects found in large-scale real-world datasets, such as Visual Genome~\cite{krishna2017visual} and VAW~\cite{pham2021learning}, ensuring the diversity and coherence of the refined prompt.

\item \emph{Spatial-condition sketch refinement} (Sect.~\ref{ssec:sketch_refine}) for refining rough sketches into high-quality and intention-aligned spatial conditions (\textbf{C2}).
Sketches created by novice users often exhibit unrealistic elements, such as uncoordinated shapes and inappropriate sizes.
This contrasts with the realistic masks typically used to guide T2I generative models, resulting in dissatisfactory outcomes, particularly for images with multiple objects.
To refine the sketch, we propose a sketch decompose-and-recompose approach:
First, the decompose stage iteratively focuses on each foreground object to generate reference images containing the object with a more realistic and accurate shape.
Second, in the recompose stage, we leverage Segment Anything Model~\cite{kirillov2023segment} to extract the refined mask for each object and recompose these masks together into one coherent mask.
The mask is further coupled with Canny edge~\cite{canny1986computational} conditioned ControlNet to stably generate images precisely matching user-selected shapes.

\end{enumerate}

We further build a prototype interface (Sect.~\ref{ssec:system}) that allows users to easily perform sketch control and prompt editing for interactive refinement (\textbf{C3}).
For sketch control, users can freely draw the initial sketch and adjust the recommended sketch produced by our decompose-and-recompose method and mask manipulation for each object.
For prompt editing, users can modify the detailed prompt for each semantic attribute.
We conduct a user study with 12 participants to compare our system with existing baselines (Sect.~\ref{sec:evaluation}) in terms of outcome satisfaction (Sect.~\ref{ssec:outcome}) and system evaluation, including system feature rating (Sect.~\ref{ssec:featurerating}), and system overall rating (Sect.~\ref{ssec:overallrating}). 
The results and user feedback reveal that \tool provides users with greater flexibility in expressing their creative intent and produces more precise results that closely align with their intentions. 
Finally, we discuss the applicability, model interpretability, and ethical considerations of \tool and similar approaches (Sect.~\ref{sec:discussion}).
The dataset and code are available at \url{https://github.com/SellLin/SketchFlex}.

Our contributions can be summarized as follows:
\begin{itemize}
    \item We introduce a novel framework that integrates sketch-aware prompt recommendation and spatial-condition sketch refinement for sketch-based image generation, to improve generation quality and ensure the alignment with user intentions, especially for novice users.
    
    \item We design a comprehensive semantic space that incorporates single-object visual attributes and multi-object relationships for generating images guided by sketches.
    The space facilitates prompt recommendation and sketch refinement, with the goal of improving spatial-semantic coherence in the generated images, especially when multiple objects are present.
    
    \item We develop a user interface that allows users to interactively refine text prompts and sketches. A user study has been conducted to evaluate the usability and effectiveness of the interactive system.
\end{itemize}

\section{Related work}\label{sec:related_wotk}

\subsection{Generative Model Assisted Creative Painting}
Recent years have witnessed the rapid development of text-to-image (T2I) generative models, exemplified by Midjourney~\cite{MID}, Stable Diffusion~\cite{rombach2022high}, DALL-E~\cite{ramesh2022hierarchical} and Imagen~\cite{saharia2022photorealistic}.
These models have revolutionized creative painting, offering rich inspiration and automation~\cite{vimpari2023adapt,ko2023large,chiou2023designing}, and enabling the public to create their own visual art without needing professional painting skills~\cite{davis2015enactive, hutson2023generative, shi2023understanding}.
The impressive capabilities of T2I models have raised concerns within the art and design communities.
% ~\cite{huang2024plantography,lawton2023tool,jiang2023ai,li2024realtimegen}.
A key concern is that the complete control exerted by generative models may diminish human contribution~\cite{jiang2023ai,boucher2024resistance}.
Existing T2I models are predominantly designed for end-to-end generation, bypassing the iterative "creation process" that enables creators to actively explore and refine their work~\cite{jiang2023ai,lawton2023tool}.
In contrast, traditional visual art creation with tactile and spatial control allows users to intuitively shape and manipulate creative projects~\cite{ko2023large,aharoni2017pigment, huang2024plantography, li2024realtimegen}.
To address these, researchers have been working towards human-AI co-creative painting, where AI acts not just as an autonomous creator but as a tool that refines or enhances the user-initiated content.
Sketch-to-image~\cite{zhang2023adding,mou2024t2i,koley2023picture} and image-to-image~\cite{rombach2022high, kodaira2023streamdiffusion} models can take user input and leverage latent consistency models~\cite{luo2023lcm,luo2023latent,lee2024streammultidiffusion} to refine and accelerate image generation with image conditioning.
In particular, some of these AI-assisted painting techniques have been integrated into industry standard tools such as Adobe Firefly~\cite{adobe} and Krita~\cite{krita}.

% lead to the key point that different type of user have different input level. rough input should be supported for novice/average user while also for expert who only have vague idea or time.
However, these condition control models require high-quality fine-grained user inputs, such as precise line drawing~\cite{zhang2023adding} or color blocking~\cite{rombach2022high}, to achieve the desired output.
When provided with rough input, the results can be unpredictable and unsatisfactory without accommodating a coarse-to-fine refinement process.
% Studies indicate significant differences in tool usage scenarios among different user groups~\cite{shi2023understanding}. 
% This is particularly relevant in Human-AI co-creative painting, where the user's skill level greatly impacts the quality of AI-generated images.
Experts who sketch well can fully leverage these models, whilst those with less painting ability often struggle to produce sketches that lead to the desired results.
% However, most previous works have focused on interaction through precise line drawing~\cite{zhang2023adding} or color blocking~\cite{rombach2022high}, making these models less user-friendly for individuals without professional training.
Our work targets novice users who may lack advanced drawing skills, but still wish to use AI-assisted tools to create their desired images.
These novice users prefer to use rough sketches to express their ideas, which differs significantly from the approach taken by experts~\cite{shi2023understanding}.
% Building on previous interaction processes, our system combines rough region sketch input with precise, fine-grained control over the output, offering an course-to-fine process. 
% The rough sketch is processed and enhanced through prompt tuning tailored to spatial conditions, while the unstable generation results from rough sketches are further refined through the system's canny-based fine-grained control.
To meet these requirements, we propose sketch-aware prompt recommendation, alongside a decompose-and-recompose strategy to refine rough sketches into precise controls.
Our approach provides an iterative coarse-to-fine process, mimicking the natural creative workflow.

\subsection{Prompt Tuning in Text-to-Image Generation}
% opening
The input text description, often termed a "prompt," plays a pivotal role in shaping the quality of images produced by T2I models. 
Therefore, designing and refining prompts is crucial for steering the model towards generating the desired images, referred to as "prompt tuning".
%prompt guidelines from existing research and community
Researchers have proposed guidelines for prompt design based on experimental studies and insights from online communities~\cite{wang2022diffusiondb}. 
These guidelines highlight key elements that contribute to generating high-quality images, including various types of prompt modifiers (\eg, subject, style, quality boosters)~\cite{oppenlaender2023prompting,oppenlaender2023taxonomy,liu2022design}. 
Auto-prompting techniques have also been proposed, using gradient descent to optimize or modify generated images by identifying enhanced or negative prompts~\cite{wang2024discrete, pryzant2023automatic, guo2023connecting}.
These methods are often paired with prompt datasets for better performance and guidance~\cite{cao2023beautifulprompt,hao2024optimizing}.
% interactive prompt refinement: prompt charm, promptmagician, propmt this,Ranni
In addition, interactive systems have been developed, allowing users to interactively align the prompts with their intentions~\cite{guo2024prompthis,brade2023promptify,wang2024promptcharm,feng2023promptmagician,promptpaint,zeng_2024_intent}. 
Some systems typically allow users to input an initial prompt, which is then refined through recommendations based on prompt datasets~\cite{wang2022diffusiondb}, followed by visualizations that enable users to perceive and adjust the prompt.
% For instance, 
% PromptCharm~\cite{wang2024promptcharm} refines the user’s prompt using promptist~\cite{hao2024optimizing} and allows further adjustments based on attention visualizations of tokens (i.e., input words). 
% PromptMagician~\cite{feng2023promptmagician} retrieves images and recommends prompts from a prompt database based on users’ input, with visualizations to evaluate these retrieved keywords. 
% Additionally, PrompTHis~\cite{guo2024prompthis} uses an innovative graph structure to visualize prompt branches, with nodes representing images and edges indicating differences in prompts.

% However, these interactive prompting systems focus solely on refining text, which is not aligned with users' input in the form of rough sketches.
Most of these interactive prompting systems primarily focus on refining text prompts. 
Recently, some efforts have been made to support direct image manipulation, such as inpainting~\cite{wang2024promptcharm,promptpaint}.
However, these works primarily optimize the visual appearance of specific regions, often neglecting the coherence across multiple regions.
% As a result, these methods are ineffective when the intended images contain multiple objects, as they fail to infer the relationships between the objects.
While LLMs have been employed to refine prompts with specific requirements such as format, style, and spatial arrangement~\cite{omost,yang2024mastering}, our experiments reveal that these approaches lack the ability to infer users' spatial control intents accurately.
To address this challenge, we integrate crowd-sourced data with fine-grained object attributes and relationships, to enhance the reasoning process, ensuring that the results are coherent and reflective of real-world scenarios.
Our system allows for direct sketch modifications, and the prompt refinement is updated automatically, enhancing the overall user experience.

% and does not allow users to interactively specify multi-modal prompts for spatial-semantic control.

% Others have developed datasets containing simple raw prompts paired with detailed, high-quality prompts, which are then used to train models that can convert initial prompts into more refined ones through fine-tuning and reinforcement learning~\cite{cao2023beautifulprompt,hao2024optimizing}.
% Leveraging the capabilities of large language models (LLMs) or Multi-modal large language models (MLLMs), some researchers prompt these language models to generate or refine prompts with specific requirements such as format,style and spatial arrangement~\cite{omost,yang2024mastering}.
% \yl{However, most automatic methods focus on text modality and do not consider users' spatial control intents.}

% Our work builds on the strengths of previous research \yl{while extending to the scenario of multi-modal prompting}. 
% We introduce a prompt template to guide prompt tuning, specifically tailored for spatially conditioned image generation.
% While we leverage MLLMs for automatic prompting, we go beyond text by incorporating user spatial input and spatial reasoning. 
% We also integrate crowd-sourced data to enhance the reasoning process, ensuring that results are coherent and reflective of real-world scenarios. 
% Unlike prior approaches that limit adjustments to prompt modifications, our system allows direct sketch modifications to influence prompt refinement, providing users with a more intuitive way to enhance their results.

\subsection{Spatial Control in Text-to-Image Generation}
%opening
% Though T2I models can generate visually appealing images from text descriptions, 
Relying solely on text-based conditioning for T2I generation models often falls short of meeting the diverse and complex needs of real-world applications, particularly when it comes to expressing precise spatial information~\cite{zhang2023adding}.
Recent reseach on spatial control of pre-trained T2I models introduces novel spatial conditional controls using canny~\cite{canny1986computational}, depth maps~\cite{lasinger2019towards}, and segmentation maps~\cite{zhou2017scene}.
% These methods involve training lightweight adapters to augment the original models. 
% While these fine-grained spatial conditions allow for precise control over the generated images, they also 
These methods, however, present challenges for users who may not have expertise in computer science or art.
For instance, preparing a canny-like sketch requires either preprocessing a reference image or possessing painting skills, 
% Creating a depth map also involves preprocessing or crafting a 3D model. 
% These tasks demand time for users to acquire the necessary skills.
which are time-consuming and require steep learning curve for novice users.
% region-based like
Alternatively, region-based spatial control allows users to specify prompts in different regions to guide the generation, which demands less in terms of spatial input from the user.
Current region-based methods can be categorized into two groups: training-based and training-free. Training-based methods~\cite{wang2024instancediffusion,li2023gligen,zheng2023layoutdiffusion} offer stronger control compared to training-free methods, but require retraining for new models that often cause style shifts due to updates in the model weights~\cite{omost}.
Training-free methods~\cite{endo2023masked,kim2023dense,bar2023multidiffusion,chen2024training,xie2023boxdiff} bypass the need for model retraining by utilizing T2I models internal mechanisms to constrain prompts to specific areas of the image. 
% These methods can be directly applied to most Stable Diffusion models with region control. 

Despite its advantages, region-based generation struggles to produce precise and consistent content within specified regions and is highly dependent on the quality of the prompt.
We leverage sketch-aware prompt recommendation with crowd-sourced datasets to generate high-quality prompts that are consistent with real-world scenarios.
In addition, we propose a decompose-and-recompose strategy that combines the advantages of high-quality spatial control and the flexibility of region-based generation.
The strategy entails decomposing a rough region sketch with multiple objects into individual objects, refining detailed spatial conditions for each object, and all spatial conditions to produce the final output.
This process not only ensures that the final output aligns with the user's intent but also maintains ease of use.

% ours
% In summary, while fine-grained control methods like ControlNet~\cite{zhang2023adding} and T2I-Adapter~\cite{mou2024t2i} offer precise spatial constraints for image generation, they pose challenges for non-expert users. 
% Conversely, region-based controls are more user-friendly but can produce unstable outputs due to variations in prompt quality and user input.
% Spatial control is crucial for accurately representing detailed scenarios and specific layouts in generated images, which is often required in real-world applications. 
% However, for novice users, mastering the necessary skills for precise spatial input can be daunting and time-consuming.
% Our system bridges these approaches by leveraging region-based inputs while stabilizes rough region using ControlNet's canny conditioning, ensuring the final output aligns with user intent while maintaining ease of use.

\section{PRELIMINARY STUDY}\label{sec:formative study}
To understand the practice of T2I generative models, we conducted in-depth interviews with eight users about their experience of using GenAI. 
We aim to identify the advantages and limitations of existing T2I generative models and identify design goals for potential improvement beyond existing methods. 

\subsection{Study Design}
\textbf{Participants}.
We recruited eight users of GenAI aged between 25 and 33 (4 females and 4 males), including four PhD students studying Digital Media or Computer Science (P1, P2, P7, P8), one UI/UX designer (P3) and three game concept designers (P4-P6).
All users have experience using popular generative models and tools like Stable Diffusion and Midjourney, with varying degree of expertise in GenAI, including four novice users with less than 6 months' usage (P1, P2, P4, P7), and four expert users with at least one-year experience and familiar with different models (P3, P5, P6, P8). 
In terms of painting expertise, three novices (P1, P7, P8) have no experience in painting, one beginner (P2) has limited knowledge about painting, and four experts (P3-P6) are proficient in painting.

\noindent
\textbf{Procedure}.
Each interview lasted 60–90 minutes. 
Initially, participants were asked to provide a self-introduction, focusing on their experience in GenAI usage, with questions like "\emph{have you used any text-to-image models like Midjourney or Stable Diffusion?}". 
Next, we explored various scenarios in which participants used or would like to use generative models, using both live demonstrations with Stable Diffusion or Midjourney 
% and explained prompt-based and spatial. 
together with explanations of prompt-based and spatial control.
Finally, we engaged in an open discussion about the advantages and limitations of the use cases presented, as well as the participants' expectations for future developments.

\subsection{Findings}

Two authors independently used open coding to analyze and code audio-recorded user feedback, generating initial codes to capture key insights from participants' experiences and expectations. 
The first author then worked with another co-author to refine and validate the themes through iterative discussions.
Based on this analysis, we distilled three primary user requirements, which are summarized as follows:

\subsubsection{Rough sketch based spatial control is preferable by novice users.}
Users with varying levels of experience exhibit different preferences in using spatial conditioning models. 
Experienced users in both GenAI and painting (P3, P6) have integrated these models into their workflow, significantly increasing their productivity. 
P6 shared his experience: "\emph{we create a line draft or use 3D modeling to generate a DepthMap, then use ControlNet for drawing}."
% He also mentioned that such workflow is widely used not only in his department, but also other companies.
Conversely, novice users in either GenAI or painting, despite being aware of spatial control features, often find these models difficult to use or insufficiently flexible to meet their expectations.
Some novice users in GenAI (e.g., P1, P4, P7) explained that these methods involve a steep learning curve, with program settings being overly complex. 
For instance, P7 noted, "\emph{I have tried using Canny, for example, but it doesn't understand my sketch very well. You need to adjust some settings, such as the guidance coefficient.}" 
Similarly, novice users in painting described a reliance on pre-existing spatial conditions found online, which limited their ability to fully realize their creative intentions. 
"\emph{I usually search images on the internet that are similar to what I want, like a person with a specific posture. Then I use the searched image to apply ControlNet. 
But in most cases, the searched image does not fully meet my expectations. 
Sometimes I can't even find an existing image that is close to my idea}," explained P8.

\begin{figure*}[t]
    \centering
    \includegraphics[width=0.99\textwidth]{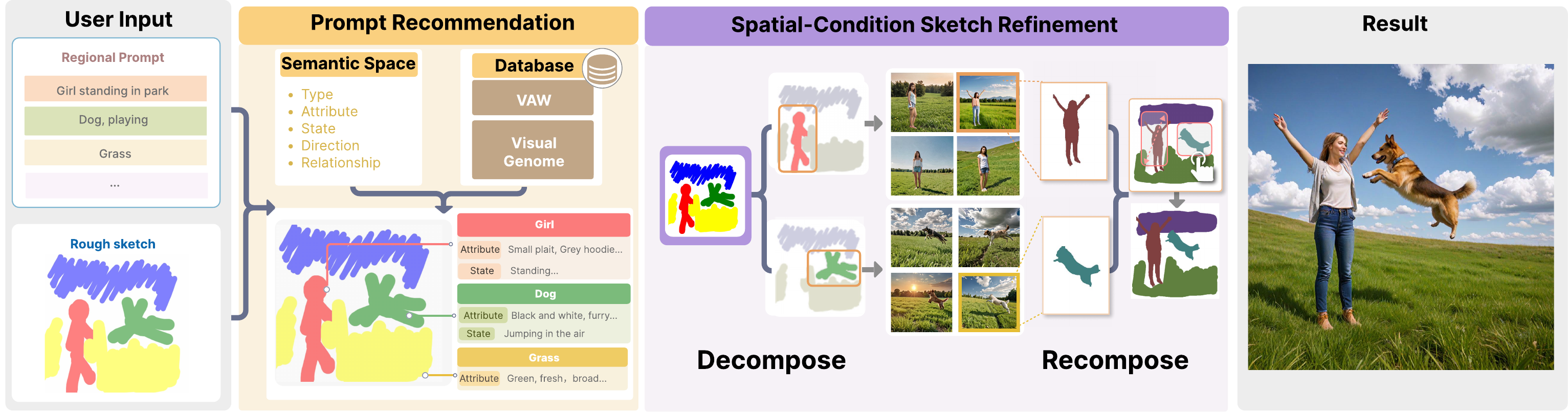}
    \vspace{-2mm}
    \caption{\tool mainly consists of three components: (1) sketch-aware prompt recommendation that support users in crafting effective prompts for the rough sketch; (2) object shape refinement through single object decomposition and generation; and (3) spatial adjustment and anchoring of object shapes.}
    \Description{\tool mainly consists of three components: (1) sketch-aware prompt recommendation that support users in crafting effective prompts for the rough sketch; (2) object shape refinement through single object decomposition and generation; and (3) spatial adjustment and anchoring of object shapes.}
    \label{fig:workflow}
    \vspace{-4mm}
\end{figure*}

Novice painters expressed strong interest in spatial control through rough sketches.
Participants with a professional painting background preferred to use line sketches and color blocks, as these techniques closely aligned with their established drawing workflows and design practices.
In contrast, novice painters prefer scribbles or regions, as these methods offer greater flexibility and require less precision when representing image elements.
P7 envisioned an interaction approach suited to their needs: "\emph{If I just draw a stick man, it knows it is a person. Then if I draw a few lines, it knows that it is an ocean. And a simple circle would represent the sun}."

\subsubsection{Combining prompt tuning and spatial control is challenging.}
Some participants familiar with spatial conditioning models reported that prompt tuning becomes more difficult given extra spatial conditions. 
This challenge arises from a potential misalignment between spatial conditions and prompts.
Such misalignment mostly occur when the intended image contains multiple objects. 
P5 found that the prompt must match the spatial condition between objects, and had to curate proper words for such prompt each time. 
P6 shared a similar point, "\emph{The count is important—if there are three people in your conditioned image but your prompt does not specify three people, the model might generate the wrong number. 
Relationships are also crucial; If you don’t specify connections, the generated result can seem disjointed, which is especially prominent in multi-diffusion}." 
The MultiDiffusion model~\cite{bar2023multidiffusion} is one of control methods that P6 frequently uses for region control.
Regarding the prompting difficulty, both P5 and P6 expressed a desire for a tool that could recommend prompts based on spatial conditions to reduce their time and cognitive load.
Despite these challenges were primarily raised by experienced participants, as novice users rarely engage with these techniques, it is clear that prompt tuning would be even more difficult for novice users.

\subsubsection{Iterative generation and refinement is difficult in end-to-end generation.}
All participants mentioned that refining end-to-end generated results is challenging. 
Specifically, they often want to change a part of the generated image while keeping other elements unchanged, as it is hard to generate image that all parts fulfill their requirement. 
However, even minor adjustments to the prompt or spatial condition can lead to significant changes in the entire generated image. 
P3 noted, "\emph{In a project, I used a reference image and made a slight change to the prompt, like adding `wearing glasses' to the description. The result looked similar, but it was just...different}."
P8 also shared, "\emph{I use AI-generated images to illustrate the scene layout for my project, including environment settings, object positioning, and content alignment. In most cases, it’s sufficient if the generated image fulfills two of these three requirements. However, in my experience, achieving all three requirements without iterating on specific areas is impossible}."
Some experienced GenAI users employ techniques like in-painting or use Photoshop for manual regional editing, but such methods require additional efforts. 
P3 explained, "\emph{I use in-painting and Photoshop to edit the parts I want to change. First, I break down the elements in the image using Photoshop, then in-paint or manually re-draw some of them. It’s time-consuming, and the in-painting doesn’t have the same coherent effect as generating the image from scratch}."
These feedback suggest that users require an iterative generation process that can progressively adjust specific parts to reach a satisfactory outcome.

%%%%%%%%%%%%%%%%%%%%%%%%%%%%%%%%%%%%%%%%%%%%%
%%%%%%%%%%%%%%%%%%%%%%%%%%%%%%%%%%%%%%%%%%%%%
\subsection{Design Goals}
The formative study illustrates the diverse methods available for spatially controlling T2I models and the needs for users in using these methods. 
However, the study also highlights that novice users often find it challenging to fine-tune prompts and prepare spatial conditions to match their intended outcomes with AI-generated images.
Our goal is to empower users with limited expertise in art and computer science to more freely create with GenAI. %using AI's generative and control capabilities. 
To achieve this, the design of a new interactive tool supporting this application should incorporate user-friendly interactions for more flexible, less demanding input and iteration while delivering high-quality results aligned with user intentions.

Based on our findings, we propose the following design goals for \tool to enhance flexible spatial control in image generation, specifically tailored for novice users.
\begin{itemize}
    \item \textbf{G1: Providing Flexible Spatial Control}.
    \tool aims to empower novice users to easily and intuitively achieve spatial control without requiring expertise in computer science or painting. 
    This control should allow users to input rough spatial conditions and provide tools for refining and enhancing the final generated result. Based on the study, we will implement region control as the rough sketch input preferred by novice users.
    \item \textbf{G2: Automating Prompt Tuning for Spatial Conditioned T2I Generation}.
    \tool should include an automated prompt tuning feature that considers both spatial conditions and the initial text prompt. 
     This can reduce the cognitive load of users while ensuring the generation of high-quality, cohesive images that accurately reflect the user's intent.
    \item \textbf{G3: Enhancing Flexibility in Image Adjustment by Decomposed Generation}.
    \tool should implement a decomposed generation approach, allowing users to make adjustments at the level of individual objects or elements within the image to facilitate iterative generation. 
    This will help avoid unintended changes to the overall image when making local adjustments in different generation rounds, thus iteratively improving user control and satisfaction.
\end{itemize}

\section{\tool System}\label{sec:system_description}

\begin{figure*}[t]
    \centering
    \includegraphics[width=0.85\textwidth]{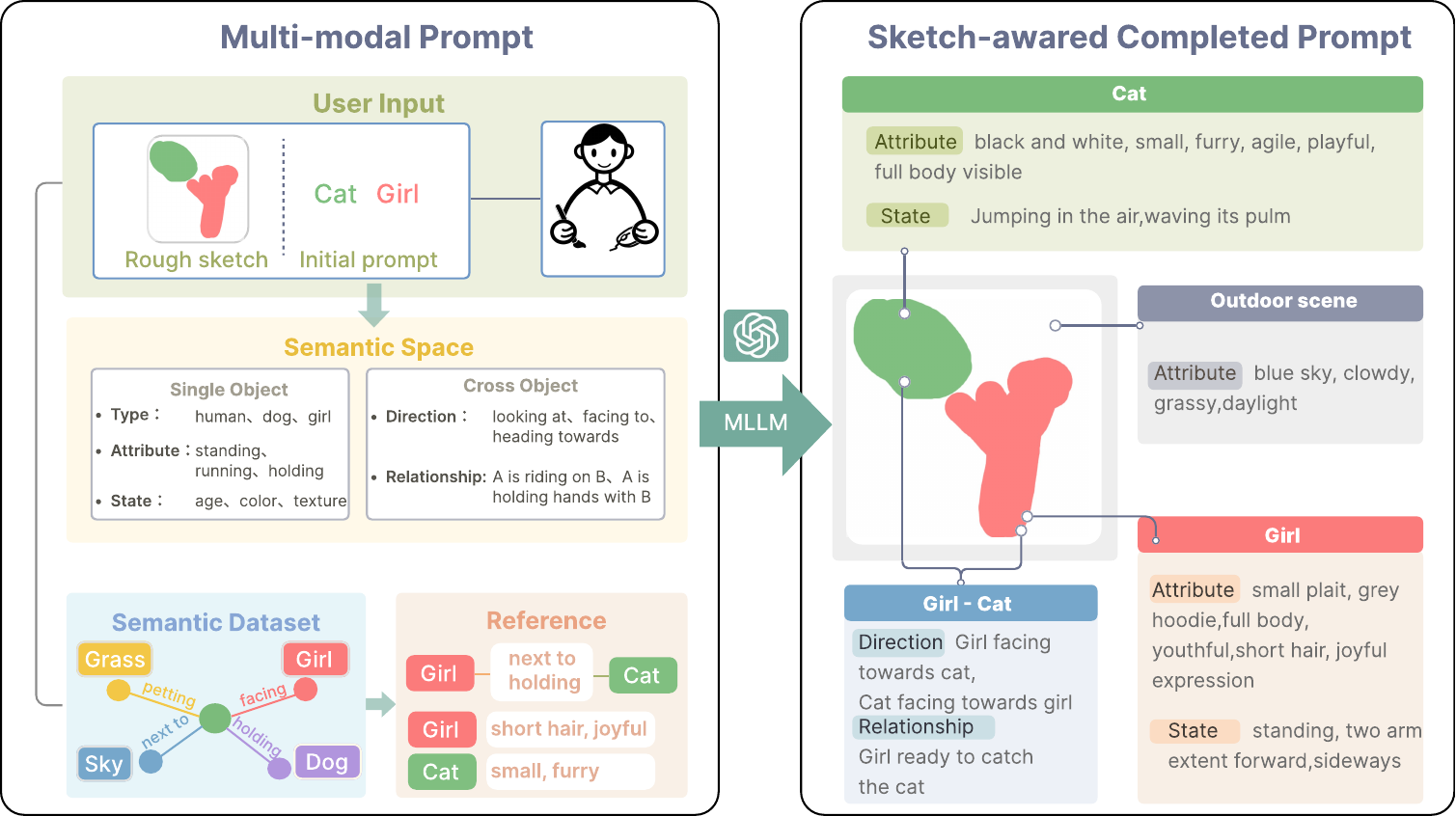}
    \vspace{-2mm}
    \caption{Our sketch-aware prompt recommendation first builds a semantic space through data-driven analysis of key semantic elements covering single object and cross object properties. Then the semantic space is integrated with retrieval of attributes and relationships reference from semantic dataset. Finally, these semantic guidance is combined with users' initial sketch to form a sketch-aware multi-modal prompt to the MLLM to support spatial-aware inference.}
    \Description{Our sketch-aware prompt recommendation first builds a semantic space through data-driven analysis of key semantic elements covering single object and cross-region properties. Then the semantic space is integrated with retrieval of attributes and relationships reference from semantic dataset. Finally, these semantic guidance is combined with users' initial sketch to form a sketch-aware multi-modal prompt to the MLLM to support spatial-aware inference.}
    \label{fig:inference}
    % \vspace{-3mm}
\end{figure*}

Based on the design goals, we design \tool, an interactive system that allows users to generate images controlled by inputs of rough sketches and simple prompts, while generating semantically cohesive and region-controlled images.
The overall framework of \tool is shown in Figure~\ref{fig:workflow}. 
The framework consists of three stages: 1) the user can draw a sketch, assign a corresponding regional prompt, and use automatic prompt recommendation to refine their initial prompt \textbf{(G1, G2)}; 2) the sketch with object decomposition and single-object generation \textbf{(G3)}; and 3) spatial adjustment and anchoring of object shapes   \textbf{(G1, G3)}.

\subsection{Sketch-Aware Prompt Recommendation}
\label{ssec:prompt_rec}
Crafting effective prompts for rough sketch-based image generation is a challenging task, as users must not only create prompts for each individual region but also ensure coherence across the entire image. 
We introduce a prompt recommendation method that automatically enhances the user’s initial input, to produce a spatially cohesive prompt that aligns with the overall composition.
Figure~\ref{fig:inference} shows the overall workflow of this process.

\subsubsection{Semantic Space Reasoning}
\begin{table}[thb] \small
    \centering
    \caption{Common examples in the semantic space across various T2I description datasets.}
    \vspace{-2mm}
    \begin{tabular}{|l|l|c|}
    \hline 
    \textbf{Space Item} & \textbf{Property} & \textbf{Instance} \\
    \hline 
    \multirow{3}{*}{\textbf{Single Object}} 
    & Type &   Man, Car, Dog, Tree, Window \\
    && Table, Ocean, Park, Wall\\ \cline{2-3}
    & Attribute & Wooden, Tall, Red, Large\\
    &&Fluffy, Round, Slim, Silver   \\ \cline{2-3}
    & State &   Standing, Moving, Swaying, \\
    &&Broken, Sleeping, Lying\\\hline
    \multirow{2}{*}{\textbf{Cross Object}} 
    &  Direction  &  Facing to, Aligned in, \\
    &&Diagonally placed\\\cline{2-3}
    & Relationship & Next to, Under, Parked on, \\
    &&Sitting by, Supporting\\ \hline
    \multirow{3}{*}{\textbf{Overall}}  
    & Lightning  &   Natrual daylight, indoor lighting, \\
    && Soft light, Moon light\\\cline{2-3}
    & Camera  &   Close-up shot, Wide-angle shot, \\
    && Overhead shot, Extreme long shot \\\cline{2-3}
    & Style  &   Realistic, Minimalist, Cinematic, \\
    && Abstractm, Anime, Oil painting\\
    \hline
    \end{tabular}
    \vspace{-1mm}
    \label{tab:example_space}
\end{table}

Previous prompt-tuning methods have primarily focused on text-to-image models, which offer limited support for refining individual region prompts and often struggle to ensure cohesiveness across the entire image.
To address this, the first step is to identify the types of prompts needed to generate a coherent image under rough sketch-based control.
Specifically, we define an "\emph{semantic space}" that provides intuitive guidance that helps users to easily input and adjust their prompts in the appropriate regions.
The process of constructing this semantic space involves analyzing online sources~\cite{sd,wang2022diffusiondb,civitai,xie2023prompt} and prompt-guideline literature~\cite{oppenlaender2023prompting,oppenlaender2023taxonomy,liu2022design} to identify critical elements that contribute to high-quality image generation. Additionally, examining image description datasets in computer vision—including traditional task datasets~\cite{caesar2018coco,pham2021learning,krishna2017visual} and recent datasets tailored for image generation~\cite{onoe2024docci}—helps uncover relevant dimensions for describing images.

\begin{figure*}[t]
    \centering
    \includegraphics[width=0.995\textwidth]{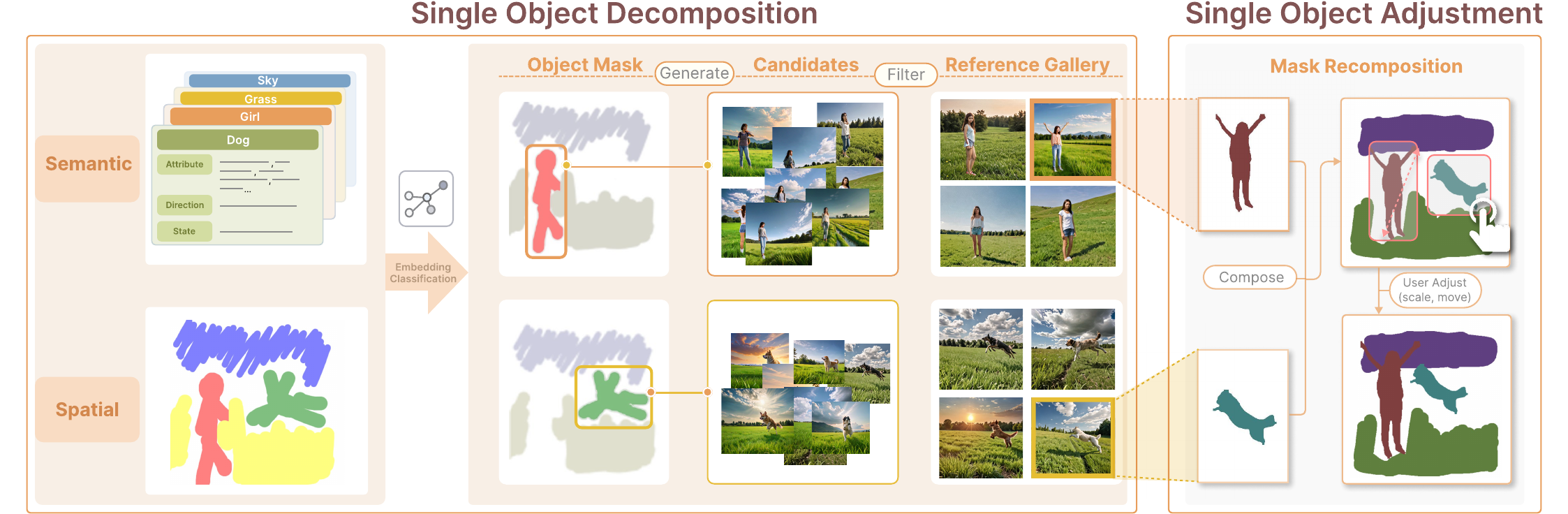}
    \vspace{-2mm}
    \caption{Spatial-condition sketch refinement can help novice users refine their sketch by generating more realistic and accurate sketch for each object through single object decomposition and generation, and subsequently allowing users to interactively refine the sketch by object selection and spatial adjustment.}
    \Description{Spatial-condition sketch refinement can help novice users refine their sketch by generating more realistic and accurate sketch for each object through single object decomposition and generation, and subsequently allowing users to interactively refine the sketch by object selection and spatial adjustment.}
    \label{fig:decompose}
    \vspace{-3mm}
\end{figure*}

\begin{table*}[thb] 
    \centering
    \caption{Example Statistics of objects, attributes and relationships in Visual Genome~\cite{krishna2017visual} and VAW~\cite{pham2021learning}.}
    \vspace{-2mm}
    \begin{tabular}{lccccc}
    \hline 
    \textbf{Space Item} & \textbf{1st} & \textbf{2nd} & \textbf{3rd} & \textbf{20th} & \textbf{50th}  \\
    \hline 
    Object & window (52k)  & man (52k) & shirt (39k) & trees (17k) & sidewalk (8k)  \\
    Attribute & white (311k) & black (195k) &  blue (118k) & clear (15k) & colorful (5.8k)   \\
    Relationship  & on (645k) & has (245k) &  in (219k) & sitting on (13k) & laying on (3.5k)    \\
    \hline
    \end{tabular}
    \vspace{-1mm}
    \label{tab:statsitic}
\end{table*}

We summarize these prompt aspects and conduct experiments to identify the key components essential for high-quality output. 
% Finally, we propose a semantic space that integrates these necessary elements to ensure coherence when generating images from rough sketches.
Table~\ref{tab:example_space} presents the common dimensions and example instances of the identified semantic space within the T2I datasets.
The specific elements within the space are listed below.
\begin{itemize}
    \item \textbf{Single Object Prompt} contains local prompts for each single object. 
    It contains \textit{type} of the object; \textit{attribute} of object including main attributes such as color, texture, shape; and \textit{state} indicates how the object acts, including still, standing, running, etc. The background is a special object that only has type and attribute.
    \item \textbf{Cross Object Prompt} explicitly specifies how objects in different regions interact, which is crucial for the coherence of the generated image.
    It includes \textit{direction} of objects and \textit{relationship} that multiple objects interact with each other.
    \item \textbf{Overall Prompt} does not directly affect multi-object cohesiveness but allows users to optionally specify the overall visual effect, including \textit{lighting}, \textit{style} and \textit{camera}.
    % \item \textbf{Overall Prompt} is not indispensable for a semantically cohesive image generation, but has significant impact in visual effect, include \textit{lighting}, \textit{style} and \textit{camera}. 
    % Previous works consider quality modifiers (e.g., best quality, high resolution). In our experiment, the state-of-the-art model can already generate high quality images without these modifiers.
\end{itemize}

Figure~\ref{fig:inference} (right) illustrates a completed semantic space for a sketch featuring a girl, a cat, and a background (\textit{i.e.}, areas without objects).
Once the individual prompt components are defined, the separate prompts within a single region are concatenated into one unified prompt using commas (\textit{e.g.}, "type: girl, attribute: long hair" becomes "girl, long hair"). 
% To ensure that the concatenated prompt remains within the 77-token limit imposed by CLIP, we employ the bag-of-conditions approach~\cite{omost}.}

\subsubsection{Sketch-Augmented Prompting}
While the semantic space simplifies the prompt input and adjustment process, manually entering all prompts and identifying the appropriate ones can still be labor intensive and cognitively demanding, particularly when dealing with many objects. 
To alleviate this burden, we utilize a MLLM GPT-4o, to automatically complete the semantic space based on the user's sketch and initial prompt.

Specifically, the user’s initial prompt and rough sketch are input into the MLLM, with the semantic space acting as a contextual guide within the prompt template. The model generates prompts to populate the semantic space, incorporating both the initial input prompt and the sketch. It utilizes spatial reasoning, guided by the Chain-of-Thought~\cite{wei2022chain} strategy, to account for the \textit{shape}, \textit{location}, and \textit{interaction} of the objects within the user's sketch.
However, relying solely on the MLLM to fill the semantic space can be risky, as it may overfit to certain content or produce results with reduced coherence~\cite{cao2023beautifulprompt}. 
To address this, we further enhance the process by retrieving reference attributes and relationships between objects from crowd-sourced text-image datasets based on real-world images~\cite{pham2021learning,krishna2017visual}. 
The datasets are organized into a dictionary, where object names serve as keys and their corresponding attributes or relationships are stored as values. 
During retrieval, \tool randomly samples $k=10$ examples from the values based on the given object names as keys, providing the MLLM with reference data. 
The raw datasets are sourced and publicly available from ~\cite{pham2021learning,krishna2017visual}.
Example statistics are shown in Table~\ref{tab:statsitic}. 
The rich semantics in these datasets enhance both diversity and coherence in the completed semantic space.

% Table~\ref{tab} presents example statistics of objects, attributes, and relationships within these datasets. 
The overall prompt generation process, as shown in Figure~\ref{fig:inference}, incorporates user input, semantic space, and retrieved attributes and relationships to guide generation. 
We employ few-shot learning~\cite{wang2023large} to enhance the quality and robustness of the generated prompts.

\begin{figure*}[t]
    \centering
    \includegraphics[width=0.995\textwidth]{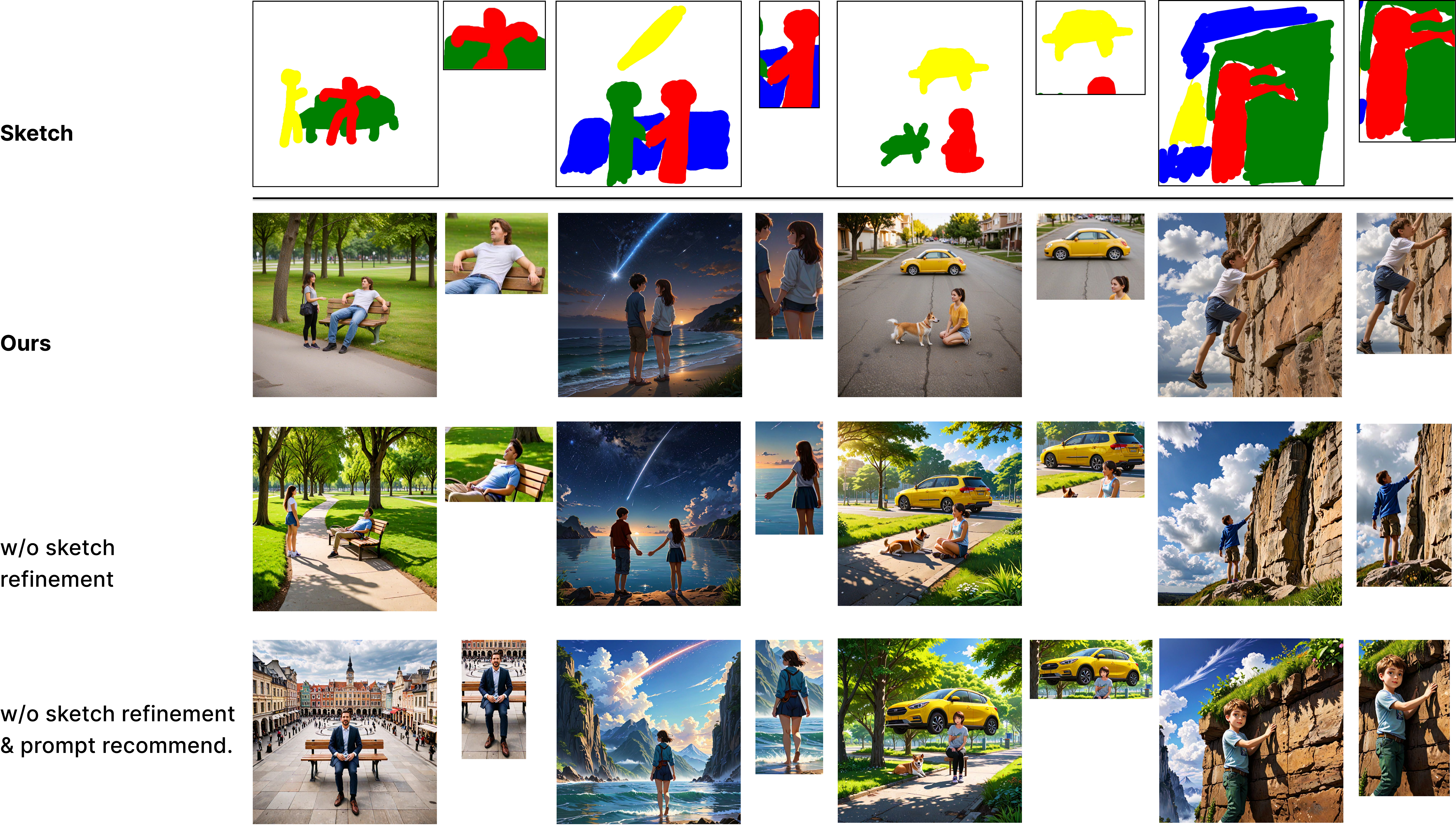}
    \vspace{-2mm}
    \caption{Ablation study shows that prompt recommendation avoids common issues like missing objects and unrealistic relationships while sketch refinement further enhances fine-grained control.}
    \Description{Ablation study shows that prompt recommendation avoids common issues like missing objects and unrealistic relationships while sketch refinement further enhances fine-grained control.}
    \label{fig:ablation}
    \vspace{-4mm}
\end{figure*}

\subsection{Spatial-Condition Sketch Refinement}
\label{ssec:sketch_refine}
To help users refine their rough sketches into fine-grained shapes that align with their intentions and allow for iterative refinement, we propose a decompose-and-recompose approach. 
As Figure~\ref{fig:decompose} shows, the sketch is first decomposed into individual objects.
The users can then generate, select, and adjust the desired single-object images. 
Finally, these selected objects, along with their spatial conditions, are combined to generate the final result.

\subsubsection{Single Object Decomposition}
Instead of directly using a single object sketch for generation, we first classify objects into two categories: \textbf{thing} as foreground object with specific shapes, such as humans, animals, or chairs, and \textbf{stuff} as background object without a defined shape, like oceans, grass, or sky, based on ~\cite{caesar2018coco}. 
% \new{"Things" are objects with specific shapes, such as humans, animals, or chairs, while "stuff" refers to objects without a defined shape, like oceans, grass, or the sky.}
To classify an object, we compute the word embedding of its type and find the nearest match in a category list containing "things" and "stuff" in the object list in the COCO-Stuff dataset~\cite{caesar2018coco}.
During single object decomposition stage, each thing object is extracted using FAST SAM (Segment Anything)~\cite{zhao2023fast}, to make sure the single object generation maximally preserves consistency to the original sketch.
Once the single object sketch is decomposed, the generation of each individual object is carried out. 
Since the goal is to generate fine-grained object shapes but not the final image, the process can be accelerated by using low-step inference (6 steps) with the Lightning Diffusion model~\cite{luo2023lcm} and a lower resolution (512x512). 
In our experiment, generating 12 images took approximately 4 seconds on a GTX 4090.

To ensure that the generated result aligns more closely with the user's sketch, we filter the generated images based on the Intersection over Union (IoU) and CLIP score~\cite{hessel2021clipscore}, which measure spatial correspondence and semantic alignment, respectively, between the user sketch and the generated single object. 
We compute a weighted sum of the IoU and CLIP scores, then sort the images and select the top four for the user to choose from.
Once the user selects an image containing the desired object shape, the target object is automatically extracted using FAST SAM~\cite{zhao2023fast}.
Each object with refined shape will automatically replace the original rough sketch.
If users have no desired image in one generation, they can perform multi-round generation until finding the desired one.
Figure~\ref{fig:decompose} shows an example in which the rough sketch of a girl and a dog is refined to specific shapes.

\begin{figure*}[t]
    \centering
    \includegraphics[width=0.8\textwidth]{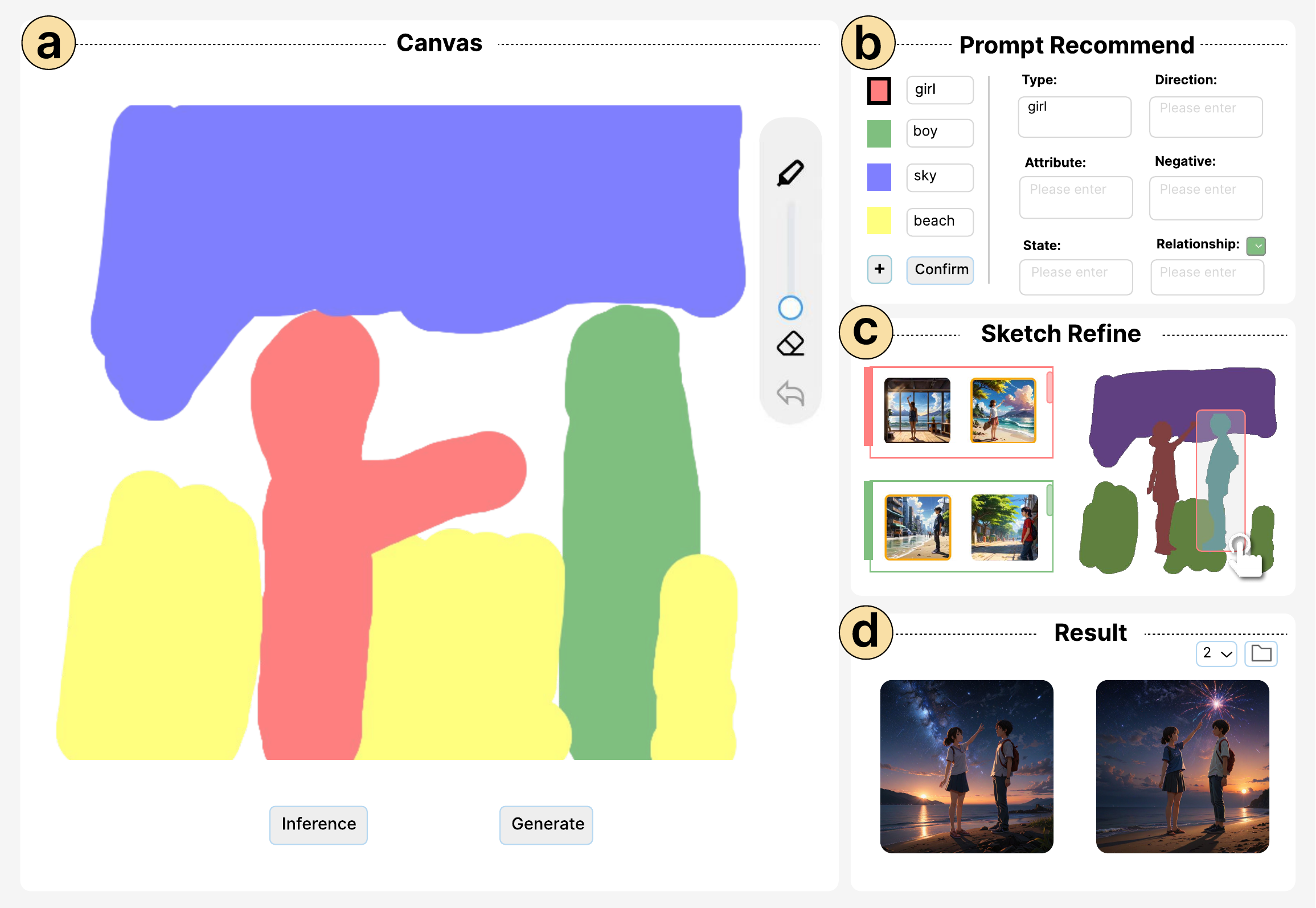}
    \vspace{-3mm}
    \caption{\tool interface consists of (a) Canvas view, (b) Prompt Recommend view, (c) Sketch Refine view and (d) Result view.}
    \Description{SketchFlex interface consists of (a) Canvas view, (b) Prompt Recommend view, (c) Sketch Refine view and (d) Result view.}
    \label{fig:interface}
    \vspace{-3mm}
\end{figure*}

\subsubsection{Single Object Adjustment}
Our system provides flexible control not only over the shapes of objects but also over their size and position. As shown in Figure~\ref{fig:decompose}
, users can easily adjust the size and spatial placement of each object. 
Once adjustments are made, the corresponding single object shape mask is moved accordingly. 
All individual object shapes are then combined into an "anchor" image.
% The term "anchor" represents the idea that the generated result, like a boat floating within the rough sketch regions, may vary with each generation, just as a boat drifts on water. 
% However, with the anchor in place, the boat remains fixed to a specific location—similar to how the selected object shapes remain fixed in the generated result. 
Shape anchoring refers to the process of fixing the object shapes in the generated result.
To apply this shape anchoring, we extract the edges of the selected object using Canny edge detection and feed them into ControlNet to ensure that the final output adheres to the user's shape preferences. 
Once users have made their desired adjustments, they can generate an image with the exact fine-grained shapes they prefer.

A related issue with the original rough sketch-based generation is that a single prompt is applied to each region separately. 
To generate images that capture relationships between objects, we create a joint mask, \textit{i.e.}, the union of object masks, for two related objects, allowing the relationship prompt to influence the interaction between them. 
Equation 1 illustrates the creation of a joint mask for the relationship between region $i$ and region $j$, 
where $\oplus$ denotes the concatenation operation.

\begin{equation}
M_{ij} = [M_i \oplus M_j].
\end{equation}

However, using a joint mask alone can sometimes result in multiple objects being generated within a single object area. 
To address this, we apply negative prompts to exclude objects outside the intended region, preventing unwanted elements from appearing in the wrong areas, as shown in Equations 2-3.
In Equation 2, the cross-attention map that correlates text and image patches is updated such that the text embedding $C_i$ is amplified by a scalar $\lambda_{m_i}$ within the masked region $M_i$.
In Equation 3, the relationship embedding condition is reinforced within $M_{ij}$, while the influence of $C_i / C_j$ is reduced in the complementary areas.

\begin{equation}
A_i \leftarrow \lambda_{m_i} \cdot C_i \odot M_i,
\end{equation}

\begin{equation}
A_{(i,j)} \leftarrow \lambda_{m_{ij}} \cdot C_{ij} \odot M_{ij} - \lambda_{m_{ij}} \cdot C_i \odot (M_{ij}-M_i) - \lambda_{m_{ij}} \cdot C_j \odot (M_{ij} - M_j).
\end{equation}

Figure~\ref{fig:ablation} shows the ablation results of our system’s functions.
Without prompt recommendation and sketch refinement, the rough sketch-based generation often produces undesirable outcomes, such as missing objects, unrealistic relationships, and incorrect perspectives.
By incorporating our prompt recommendation, which refines prompts to be spatially aligned with the sketch, the results become more stable and coherent, with the generated objects closely matching the original sketch.
Finally, with prompt recommendation and sketch refinement, users can achieve fine-grained control, enabling them to generate detailed results that align with their sketches and creative intentions. For example, in the first column, the posture of the man lying on the chair with both arms spread out is more accurately captured in our result compared to the other two. 
In the third column, the girl's head is positioned slightly to the right and below the car within the mask, which aligns with our result. In contrast, in the other two results, the girl's head overlaps with the car, resulting in an incorrect spatial relationship.

\subsection{System Interface}\label{ssec:system}
The interface of \tool is built with a back-end based on Python and a front-end based on Vue.js with additional Fabric.js-supported canvas interactions.  
As shown in Figure~\ref{fig:interface}, the \tool system consists of four main views: Canvas view, Prompt Recommend view, Sketch Refine view, and Result view.

\textbf{Canvas view}.
The Canvas view (Figure~\ref{fig:interface} (a)) is the main playground of our system where users interactively specify initial spatial control.
Its main body is a canvas on which users can freely draw rough sketches of different colors corresponding to different types of objects. 
On the right side is a series of control buttons that allow users to perform basic brush control such as changing stroke width, switching to eraser and undoing previous stroke.

\textbf{Prompt Recommend view}. 
As shown in Figure~\ref{fig:interface} (b), in this view, users can input sophisticated prompt based on our semantic space to control the sketch of each individual object or background.
They can choose from manual input or click the \emph{Inference} button upon finishing the sketch on Canvas view to leverage our sketch-aware prompt recommendation to automatically suggest the appropriate prompt for each dimension.

\textbf{Sketch Refine view}.
This view enables users to utilize our spatial-conditioned sketch refinement method to refine their initial sketch by selecting a more realistic and precise recommended sketch for each object.
The refined sketch is displayed on a canvas on the right, where users can further adjust the position and size.
For example, as shown in Figure~\ref{fig:interface} (c), the users refine the sketches for both \emph{girl} and \emph{boy}.

\textbf{Result view}.
Finally, the Result view (Figure~\ref{fig:interface} (d)) displays the generated images. 
Users can choose the number of generated samples to show.
They can also save their favorite results.

% On the right is semantic input panel, where users can provide fine-grained multi-faceted semantic control on both global and individual mask level based on our design space.
% Particularly, users can choose to either manually enter the semantics on different aspects or leverage our retrieval augmented spatial-aware prompt inference to automatically suggest appropriate words to fill in each entry.

% \textbf{Image Gallery}.
% The image gallery is designed to display pre-generated single object images as well as final result image.
% For the single-object images, we display four recommended samples on the top, where users can select appropriate mask for single object.
% When a particular image is selected, the single-object mask extracted from this image will be applied to the original user-created spatial sketch, replacing the previous mask of the corresponding object.
% The adjusted spatial masks will be displayed below.
% Users can directly drag the masks to perform further modification including scaling and moving to create a more satisfactory layout.
% At the bottom of the Image Gallery, we display the final generated result.
% Users can choose the number of generated samples to show.

\section{Evaluation}\label{sec:evaluation}
To evaluate the effectiveness of \tool, we conducted a user study comparing \tool with two baseline systems: a text-to-image (T2I) and a region-to-image (R2I) generation system. 
We focus on how well our system supports novice users in creating images with minimal input while offering greater control and higher-quality outcomes compared to existing methods. 
Specifically, we assess: 1) quantitative comparison of user performance between baseline systems and \tool,  2) user satisfaction with the generated images compared to baseline systems, 3) the effectiveness of the system’s features, and 4) the overall usefulness of the system.

\subsection{Experiment Setup}

\subsubsection{Participants}
We recruited 12 participants (7 males, 5 females), aged from 18 to 30, most of whom were postgraduate students from a research university with diverse backgrounds in art, design, and computer science. 
In terms of experience in GenAI, 9 participants were novice users with less than one year or no experience, 2 participants had more than one year of experience, and 1 participant had over a year of experience. 
Regarding painting skills, 5 participants had no formal training, 5 had basic painting skills, and 2 were proficient. All participants were invited to take part in the user study in person.

\subsubsection{Baseline Models}
Since our focus is on novice users, we choose baseline image generative models that are easy to use and widely adopted by beginners. 
We select two methods: T2I generation and R2I generation, both of which allow hands-on exploration without requiring advanced expertise:
\begin{itemize}
    \item \textbf{T2I Generation} serves as a foundational approach, where users generate images solely based on text input. It is considered a baseline due to its simplicity and widespread use among novices. 
    \item \textbf{R2I Generation} allows users to specify prompts for particular areas within the canvas, providing more control over the final image. 
    For this purpose, we employed a state-of-the-art region-based generative model, with a Dense Diffusion method~\cite{kim2023dense,omost}, which offers advanced control while remaining accessible to novice users.
\end{itemize}

\subsubsection{System Implementation}
To ensure a fair comparison, all image generative models, including those in our system and the baseline systems, were build on the same backbone: the $ColorfulXL-Lightning$ model~\cite{colorful}. 
This model is a fine-tuned version of the SDXL model, optimized for generating images with high aesthetic scores, ensuring consistency in prompt interpretation, image quality, and style across all comparisons.
% \new{All three models were implemented as local web applications, enabling users to interact directly with them on a browser.}
For T2I generation, users were asked to use the Stable Diffusion WebUI~\cite{sd}, which is the most widely used local web platform.
For both the R2I generation system and \tool, they are implemented within the same web application, as illustrated in Figure~\ref{fig:interface}. 
Both the R2I system and \tool allow users to sketch in the same way; however, in the R2I generation condition, the advanced features of \tool are disabled. 
Specifically, in the R2I condition, users can only access a sketch board and a single text box to specify prompts for the selected region, 
without the sketch-aware prompt recommendation and spatial-condition sketch refinement.
% rather than using separate semantic spaces (e.g., type, attribute, etc.), which disabled text boxes are made invisible on the front end.

% To further assist users, they were allowed to use ChatGPT, Omost, and Promptist to generate prompts when working with the baseline models. 

\subsubsection{Procedure}
Each participant followed the procedure as outlined below. Users were allowed to take a break after each task.

\begin{itemize}
    \item \textbf{Introduction}.
    We began by collecting demographic information from the participants and obtaining their consent to participate in the study and allow for the collection and analysis of the result data. 
    Following this, we introduced both our system and the baseline systems, explaining their functionality. 
    Participants were then given five minutes to freely explore and familiarize themselves with each system.

    \item \textbf{Close-Ended Task}.
    We carefully designed two image generation tasks. 
    The first task involved generating a simple image featuring two main objects, while the second, more complex task required creating an image with four main objects, with more intricate relationships between them. 
    For each task, we provided a reference image and participants were asked to use both the baseline systems and our system to generate images as similar as possible to the given reference image.
    To accommodate different creative preferences, we offered three style options: realistic, animated, and painting, allowing participants to freely choose their preferred style for each task. 
    % The style-related prompts were pre-written, enabling participants to focus solely on the content of the image.
    Task assignment order was randomized to mitigate learning effects between tasks.
    The tool assignment order was fixed to T2I, R2I, and \tool. 
    % This order ensures that there is no explicit prior knowledge leakage between tools. 
    This order ensures that there is no explicit prior knowledge leakage from advanced tool to baselines. 
    For instance, using R2I after \tool could result in participants already knowing the recommended prompt from \tool when they use the R2I system, which should not be available to users in the baselines.

\begin{figure*}[t]
    \centering
    \includegraphics[width=0.9\textwidth]{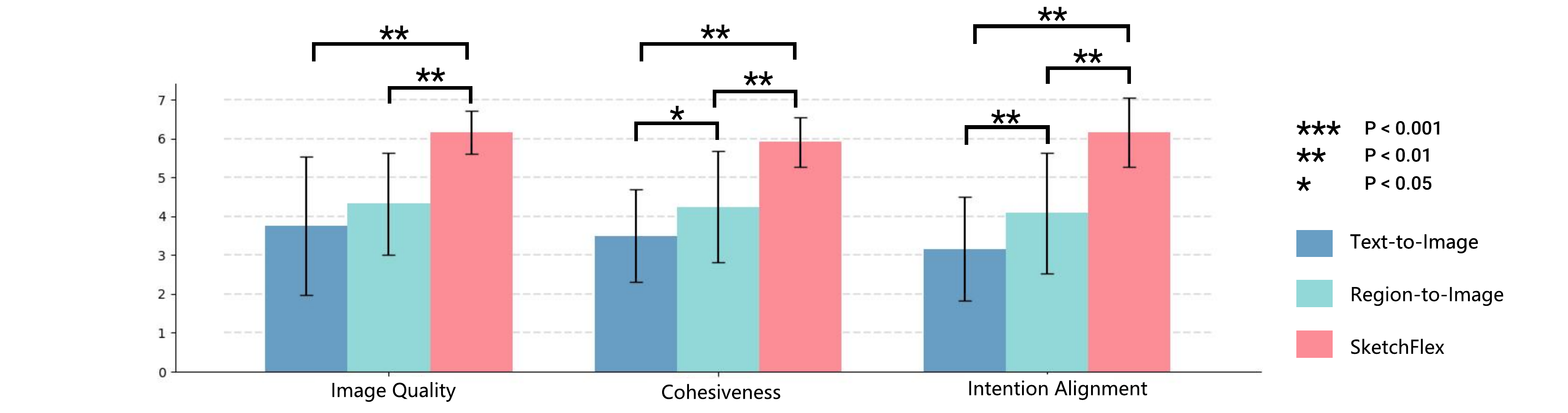}
    \vspace{-2mm}
    \caption{Outcome satisfaction survey indicates that \tool significantly outperforms the baseline text-to-image model and region-to-image model in image quality, cohesiveness and intention alignment.}
    \Description{Outcome satisfaction survey indicates that \tool significantly outperforms the baseline text-to-image model and region-to-image model in image quality, cohesiveness and intention alignment.}
    \label{fig:outcome}
    \vspace{-3mm}
\end{figure*}
    
    \item \textbf{Open-Ended Task}.
    The open-ended task allowed users to freely create their own work using our system without any constraints. 
    They were encouraged to explore and experiment, using their own ideas throughout the process.
    To ensure the tasks accurately reflected real-world scenarios and iterative exploration, images in both the close-ended and open-ended tasks were generated using random seeds. 
    The random seeds enable users to iteratively explore different results, simulating real-world conditions of free generation.
    % where users typically do not have a fixed 'seed' for the images they aim to generate.   
    
    \item \textbf{Survey and Interview}.
    After completing the tasks, participants were asked to fill out a 7-point Likert scale questionnaire to assess the system's outcome satisfaction, feature usability, and overall performance. 
    For system features, we focused on prompt recommendation, single object decomposition, and single object adjustment. 
    In terms of overall performance, participants rated the system based on its usefulness, flexibility, controllability, and engagement. 
    Following the questionnaire, we conducted a semi-structured interview with each participant to gather detailed feedback on their experience using our system. 
    For brevity, we use parentheses (e.g., 5/12) to indicate the number of users who agreed or disagreed during the interview.

\end{itemize}

\subsection{Quantitative Evaluation of User Performance}

We evaluated user performance on close-ended tasks using Intersection over Union (IoU), which measures the area alignment between objects in reference images and user-generated results. 
To segment object masks (e.g., girl, cat in Task 1 and girl, boy, train in Task 2), we employed Segment Anything (SAM)~\cite{kirillov2023segment} for open-set segmentation and Dino~\cite{liu2023grounding} for semantic guidance. 
A higher IoU score indicates better alignment and superior performance.

\begin{table}[ht] \small
    \centering
    \caption{User performance in close-ended task measured by IoU.}
    \vspace{-2mm}
    \begin{tabular}{lcccc|cc}
    \hline 
    \textbf{Methods} & \multicolumn{2}{c}{\textbf{Task 1}} & \multicolumn{2}{c}{\textbf{Task 2}} & \multicolumn{2}{c}{\textbf{Overall}} \\
    & \textbf{mean} & \textbf{SD} & \textbf{mean} & \textbf{SD} & \textbf{mean} & \textbf{SD} \\
    \hline 
    Text-to-Image (T2I) & 0.312 & 0.115 & 0.316 & 0.083 & 0.314 & 0.100 \\
    Region-to-Image (R2I) & 0.440 & 0.105 & 0.396 & 0.099 & 0.418 & 0.102 \\
    SketchFlex  & \textbf{0.613} & 0.073 & \textbf{0.456} & 0.107 & \textbf{0.535} & 0.092 \\
    \hline
    \end{tabular}
    % \vspace{-3mm}
    \label{tab:IoU}
\end{table}

The results are shown in Table~\ref{tab:IoU}.
Overall, \tool achieved an IoU score of 0.535, outperforming the T2I method (0.314) and the R2I method (0.418). 
We first performed Kruskal-Wallis tests to evaluate the significant difference among the three methods and Welch’s t-tests to measure the significant difference between each two methods.
Tasks 1 and 2 produced Kruskal-Wallis test results with $p<0.001$ and $p<0.05$.
In task 1, \tool achieved a score of 0.613, significantly outperforming T2I (0.312, $p<0.001$) and R2I (0.440, $p<0.001$). 
However, in Task 2, the performance gap narrows, with \tool scoring 0.456 compared to T2I (0.316, $p<0.01$) and R2I (0.396, $p = 0.19$).

Task complexity appears to play a key role in these trends. 
Task 1 involves fewer objects (two) with larger areas, making it easier for users to adjust object shapes and sizes. 
In contrast, task 2 features three smaller objects, requiring more precise arrangement to match the reference image, which increases difficulty. 
This is further reflected in the standard deviation (SD) values. 
In task 1, \tool has a lower SD (0.073), indicating more consistent performance, while in Task 2, its SD increases to 0.107—slightly higher than T2I (0.083) and R2I (0.099). 
By analyzing the result, we find that firstly the flexibility of \tool allows users to make more subjective decisions about how to align masks with the reference image, and such subjective gap is enlarged in more complex task. 
For instance, some users focus on the size of objects while overlooking their exact positions, or vice versa, resulting in inconsistent outcomes across different users.
Also, while skilled users of painting can leverage this flexibility to achieve highly accurate results, less experienced users may struggle with detailed manipulations, leading to greater variability in performance. 
We further discuss the different performance for users with different skill levels in Section~\ref{sec:nVe}.
Finally, while the R2I method (0.418) performs better than T2I (0.314), its improvement is more modest compared to \tool, as loosely defined regions lack consistent size, shape, and position, limiting its precision and overall effectiveness.

\subsection{Outcome Satisfaction}\label{ssec:outcome}

\subsubsection{Quantitative Comparison}
% \subsubsection{Outcome Satisfaction}

    We compared the results generated from our system with those produced by the two baselines in close-ended tasks. 
    Outcome satisfaction was evaluated across three key dimensions: \textbf{Intention Alignment}, which measures how well the generated result aligns with the user's intended outcome; \textbf{Cohesiveness}, which assesses whether the generated image includes all objects specified in the prompt and whether they are depicted in a natural state with coherent relationships; and \textbf{Image Quality}, which evaluates the overall quality of the image, including resolution, color vibrancy, visual appeal, and absence of distortion.

    To determine significant differences between approaches, we first performed Kruskal-Wallis tests, followed by pairwise Wilcoxon signed-rank tests with Bonferroni correction for p-values ($\alpha = 0.05$). Significant values are reported at the levels of $p<0.05 (*)$, $p<0.01 (**)$, and $p<0.001 (***)$. 
    All three evaluated dimensions yielded Kruskal-Wallis test results with $p<0.001$.
    Figure~\ref{fig:outcome} presents user ratings in terms of outcome satisfaction.

\begin{figure*}[t]
    \centering
    \includegraphics[width=0.95\textwidth]{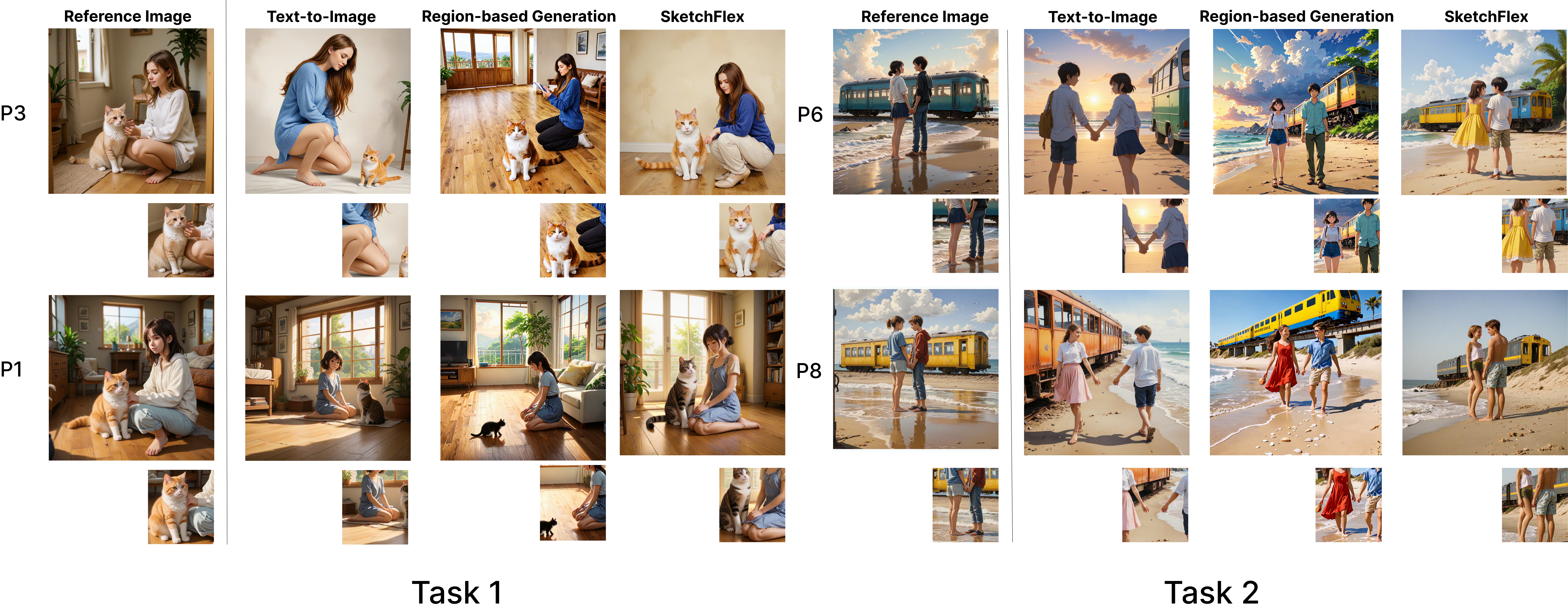}
    \vspace{-2mm}
    \caption{Outcome examples of Task 1 and Task2 show that while region-based generation offers better spatial control than Text-to-Image, it can not perform precise control, \tool can help users more precisely replicate the spatial composition of the given reference than other two baselines.}
    \Description{Outcome examples of Task 1 and Task2 show that while region-based generation offers better spatial control than Text-to-Image, it can not perform precise control, \tool can help users more precisely replicate the spatial composition of the given reference than other two baselines.}
    \label{fig:outcome_task1}
    \vspace{-1mm}
\end{figure*}

\begin{itemize}
    \item \textbf{Image Quality}.
    \tool’s outcomes (mean = 6.17, SD = 0.55) significantly outperformed both the text-to-image method (mean = 3.75, SD = 1.78, $p$ < 0.01) and the region-based image generation method (mean = 4.33, SD = 1.31, $p$ < 0.01) in terms of image quality. 
    While all participants agreed that the image quality produced by \tool was superior, most (8/12) felt that the gap in image quality between the methods was not substantial. 
    This can be attributed to the backbone model, which is fine-tuned to generate images with high aesthetic scores, demonstrating strong generative capabilities across different conditions.
    Additionally, half of the participants (7/12) reported that region-based generation with initial prompts sometimes resulted in lower image quality compared to text-to-image generation under similar prompt settings. 
    For instance, P6 mentioned, "\emph{I wrote a very simple prompt for the colored regions and didn’t write a prompt for the background, and the generated result ended up being just some decorative patterns without any meaningful content}."
    This occurs because prompts are more critical in region-based models, and when certain areas are left without specific prompts, it can result in noticeable defects in image quality. 
    This observation partially explains why there was no significant gap between the text-to-image and region-based image generation methods ($p$ = 0.159).

    \item \textbf{Cohesiveness}.
    The results show that \tool generates more cohesive images (mean = 5.92, SD = 0.64) compared to both text-to-image (mean = 3.50, SD = 1.19, $p<0.01$) and region-based image generation (mean = 4.25, SD = 1.42, $p<0.01$).
    Half of the participants (7/12) reported that the low cohesiveness in images generated by the text-to-image method was due to misunderstandings about the number of objects or the mixing of attributes between different objects,
    while region-based generation improved this issue through region control($p<0.05$).
    A few participants (3/12) reported that the improvement of outcome in cohesiveness by \tool compared to region-based generation was largely due to the prompt recommendation system, which assigns coherent prompts to specific regions.
    As P3 noted, "\emph{When I only used simple prompts for the region-based generation, it often missed some objects. But once the system filled in these prompts, the entire image looked much better, with all the objects appearing in the right place}" (P3, Figure~\ref{fig:outcome_task1}).

    \item \textbf{Intention Alignment}.
    Participants gave significantly higher ratings for \tool in generating intention-aligned images (mean = 6.16, SD = 0.89) compared to the text-to-image method (mean = 3.16, SD = 1.34, $p<0.01$) and the region-based image generation method (mean = 4.08, SD = 1.55, $p<0.01$). 
    All participants noted that the precise shape fixing and spatial adjustments made it much easier to create the desired image.
    P9 remarked, "\emph{I can easily adjust the location and relationships between objects, and it closely matches the image in my mind after I make adjustments}." P7 also noted, "\emph{The system allows me to first fix the content of the image in my mind... The interpretation of text varies, but this visually maps what I am thinking}."
    Additionally, two participants mentioned that the improved intention alignment was partly due to the prompt recommendation feature, which provided suitable prompts corresponding to their sketches, particularly for state-related prompts they may not have considered on their own.
\end{itemize}

\begin{figure*}[t]
    \centering
    \includegraphics[width=0.95\textwidth]{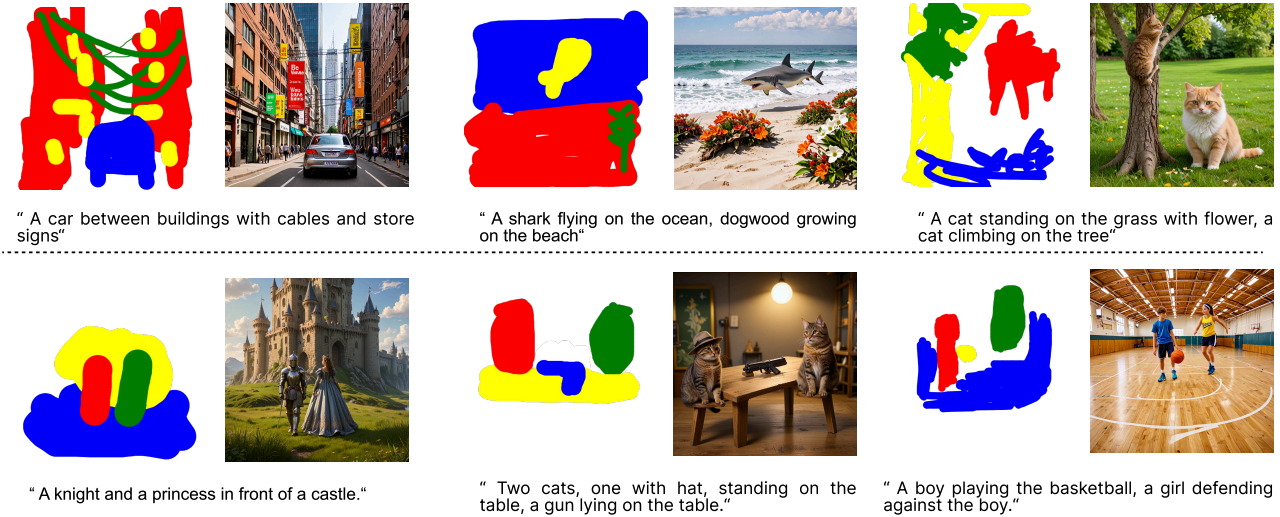}
    \vspace{-3mm}
    \caption{Outcome examples of open-ended task show that users can freely create sophisticated images with varying degree of complexity.}
    \Description{Outcome examples of open-ended task show that users can freely create sophisticated images with varying degree of complexity.}
    \label{fig:outcome_task2}
    % \vspace{-3mm}
\end{figure*}

\begin{figure*}[t]
    \centering
    \includegraphics[width=0.9\textwidth]{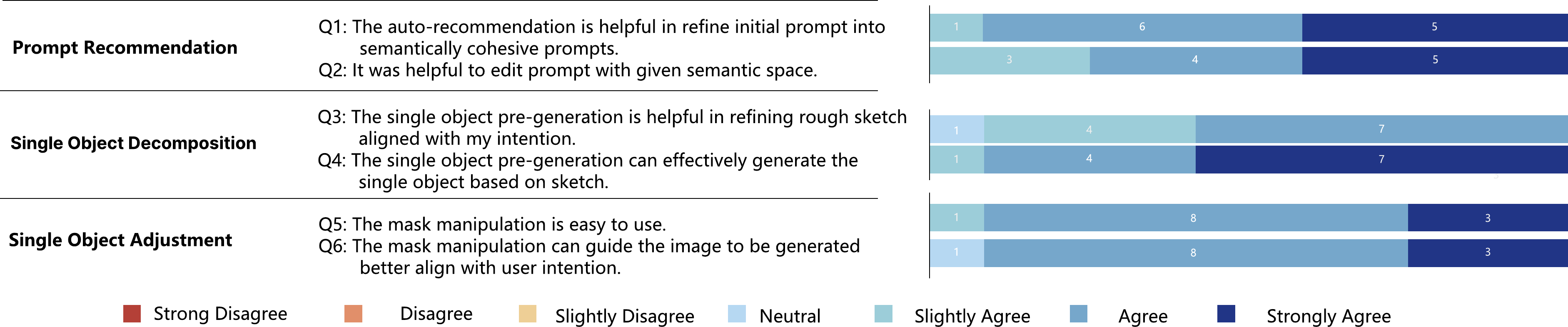}
    \vspace{-2mm}
    \caption{User ratings of main features of \tool, including Prompt Recommendation, Single Object Decomposition and Single Object Adjustment.}
    \Description{User ratings of main features of \tool, including Prompt Recommendation, Single Object Decomposition and Single Object Adjustment.}
    \label{fig:feature}
    % \vspace{-3mm}
\end{figure*}

\subsubsection{Qualitative Examples and feedback}
Figure ~\ref{fig:outcome_task1} shows examples of user-generated outcomes in Task 1 and Task 2, showcasing both realistic and anime styles.
For task 1, \tool generates a result that closely resembles the reference image, with the cat positioned in the left half and the girl's kneeling posture and hand occupying the right half. In contrast, the other results exhibit issues such as incorrect size or position.
For task 2, \tool accurately depicts two people holding hands while slightly facing each other. In comparison, the other results either fail to show them holding hands or display incorrect orientation or position.
Participants (5/12) reported that while using text to generate images can achieve similar postures, attributes, and actions, it often struggles to position objects correctly. 
Region-based generation offers better control over object location and size, but participants found it challenging to control the precise shape and state of objects with rough region sketches (9/12).
P3, P6, P7, and P10 mentioned the difficulty in positioning the train horizontally across the couple in the scene, but they found it much easier with our system.
P7 and P10 commented that the system greatly simplified this task.

With \tool, participants were able to generate images with objects in the exact shape, location, and relationships they intended, making the outcomes much more aligned with the reference images. 
P1 remarked, "\emph{I can choose the desired object shape and fix it in the generated result, like the girl and cat. The exact shape, posture, and direction are the most significant improvements compared to other methods. The system combines both text and region controls to provide shape fixing and adjustment, which is incredibly useful}" (P1, Figure~\ref{fig:outcome_task1}).
Figure ~\ref{fig:outcome_task2} illustrates examples of creative drawings produced by users with our system. Participants were able to create high-quality images that aligned with their rough region sketches. P2 commented, "The system can convert my simple input into an image, a highly complete one. It makes me feel like I can create much more freely."
% Some participants also noted that the system offered pleasant surprises and helped them explore more possibilities. P8 observed, "\emph{I initially drew a simple sketch and just thought the object should be roughly in the right position, but not exactly what it should look like. The system gave me options, and I realized that what it generated was actually what I wanted, even though I hadn’t expected it}" (P8, Figure~\ref{fig:outcome_task2}).

\begin{figure*}[t]
    \centering
    \includegraphics[width=0.9\textwidth]{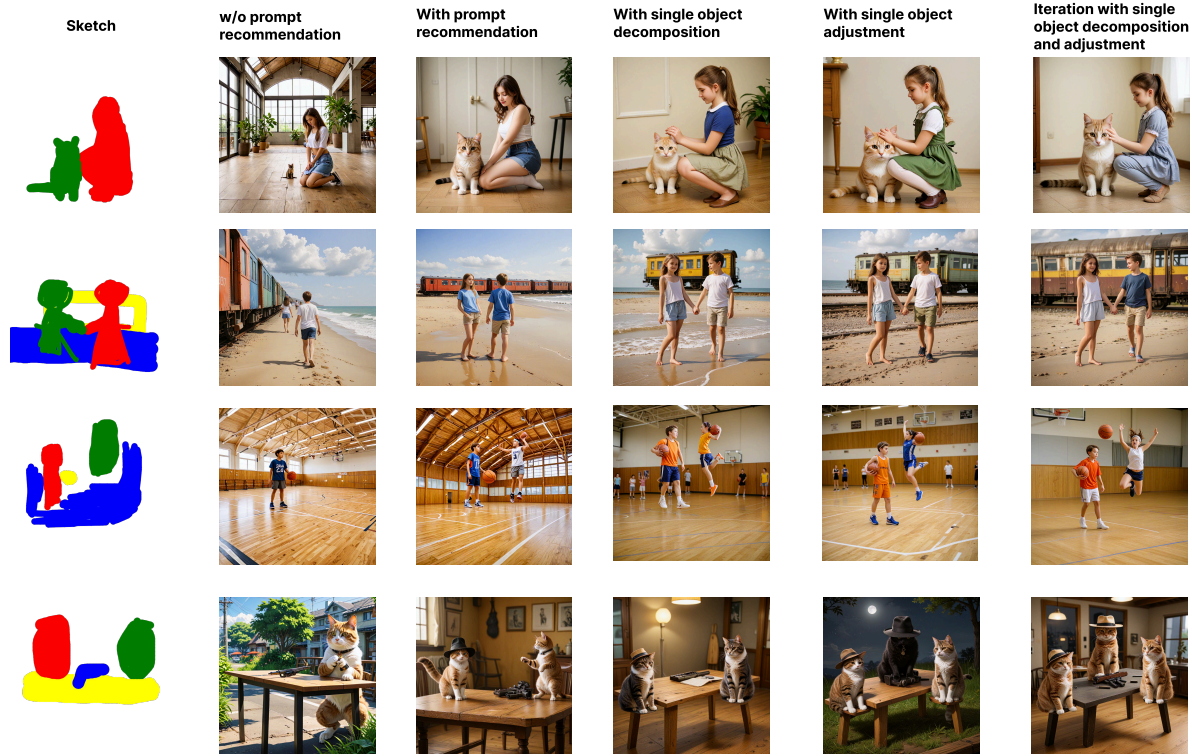}
    \vspace{-2mm}
    \caption{Examples showing single feature influence on the results, including Prompt Recommendation, Single Object Decomposition, and Single Object Adjustment.}
    \Description{Examples showing single feature influence on the results, including Prompt Recommendation, Single Object Decomposition, and Single Object Adjustment.}
    \label{fig:feature_ablation}
    \vspace{-1mm}
\end{figure*}

\subsection{System Evaluation}
\subsubsection{System Feature Rating}\label{ssec:featurerating}
We report users' ratings of our system's different features: \textit{Prompt Recommendation}, \textit{Single Object Decomposition}, and \textit{Single Object Adjustment}.
The Prompt Recommendation component includes semantic space guidance for user input as well as automatic prompt recommendation. The Single Object Decomposition module handles the decomposition and refinement of individual objects, while the Single Object Adjustment enables users to adjust the location and size of each object's shape.
Overall, all features were well-received by participants. 
The results are shown in Figure~\ref{fig:feature}.

\textbf{Prompt Recommendation}. Most participants (9/12) reported that the system significantly reduced the time spent crafting detailed prompts for each region without compromising image quality (Q1, mean: 6.08, SD: 0.75). 
P11 commented, "\emph{It quickly helps me fill in the prompts. When there are many regions, I just don’t know what the appropriate words are}."
In addition, all participants agreed that the semantic space made it easier to adjust the prompts (Q2, mean: 6.16, SD: 0.55). 
P7 explained, "\emph{If the generated image has something I don't want, I know where to find the prompt and quickly change it. If there were only a text box, it would be time-consuming to find the right one to modify}."
However, a few participants mentioned that the prompt recommendation can sometimes cause frustration when recommended prompts didn’t match their expectations. 
P2 stated, "\emph{It recommends many prompts at once, and if there are a lot I want to change, I have to rewrite them one by one}."

\textbf{Single Object Decomposition}. 
The single object decomposition feature was the most appreciated aspect of \tool. 
Participants found it highly beneficial in refining rough sketches into their desired fine-grained shapes, helping to reduce cognitive load while aligning the generated result with their intentions (Q3 \& Q4, mean: 6.5, SD: 0.64 \& mean: 5.5, SD: 0.64). 
P11 noted, "\emph{The system can quickly generate options for your rough sketch. If you don’t like it, you can keep iterating until you find one that suits you}."

\textbf{Single Object Adjustment}. 
All participants agreed that the single object adjustment was intuitive and efficiently guided image generation. 
Some participants (4/12) particularly praised its ease of use and simplicity (Q5, mean: 6.1, SD: 0.79). 
Regarding the controllability of the shapes users chose and adjusted, participants reported that it largely generated shapes as intended (Q6, mean: 6.3, SD: 0.62).
Additionally, a few participants applauded the flexibility of the system, noting that the shapes weren’t fully fixed, leaving room for subtle variations in the generation. 
P4 shared, "\emph{I positioned the boy and girl closer together and wrote a prompt for them to hold hands. Initially, they were just standing, but when the image was generated, they were actually holding hands. It was amazing}."

\subsubsection{System Feature Influence on Results}
This section reports how individual feature of \tool affects generated results in the user study, with examples illustrated in Figure~\ref{fig:feature_ablation}.

\textbf{Prompt Recommendation}. Prompt recommendation plays a crucial role in enhancing the overall cohesiveness of generated images. 
As illustrated in the second column of Figure~\ref{fig:feature_ablation}, when participants create images using their sketches and self-written prompts, several issues arise. 
These include \textit{missing objects} in [column 2 - rows 3, 4], \textit{misaligned sizes} in [column 2 - rows 1, 2, 4], and \textit{lack of natural interaction} in [column 2 - rows 1, 2].
By using the recommended prompts, as shown in column 3, these issues are significantly mitigated. 
A majority of participants (8/12) acknowledged encountering issues such as missing objects or unnatural object interaction/perception, and noted that the recommended prompts alleviated these problems. 
As P1 reported, "\emph{[with prompt recommendation], the generated image is definitely more aligned with my sketch and looks more harmonious.}"

\textbf{Single Object Decomposition.} Our decomposed strategy for single object generation significantly enhances the alignment between the generated object shapes and user preferences by providing fine-grained shape candidates and enabling iterative adjustments, as shown in column 4.
For example, in row 1, the user generated and selected a little girl with her hands slightly lifted to touch the cat, instead of the original depiction of a mature girl with her hand resting on her leg. 
This action of touching the cat more closely aligns with reference image 1 in Figure~\ref{fig:outcome_task1}. 
In row 3, the user generated and selected a girl in the green area of the sketch with a jumping posture that is more expressive and aligns better with their preferences. 
These chosen object shapes are then fixed in subsequent generations to preserve the desired object features while allowing other elements of the image to change, as demonstrated in columns 5 and 6.
Additionally, some participants mentioned that single object generation helps them explore potential shapes when they have no clear idea in mind. 
For instance, P6 remarked, "\emph{The black cat surprisingly fit the scenario that I want. It is not exactly what I'm looking for at first, but when it was generated, I knew it was the right one.}"

\begin{figure*}[t]
    \centering
    \includegraphics[width=0.82\textwidth]{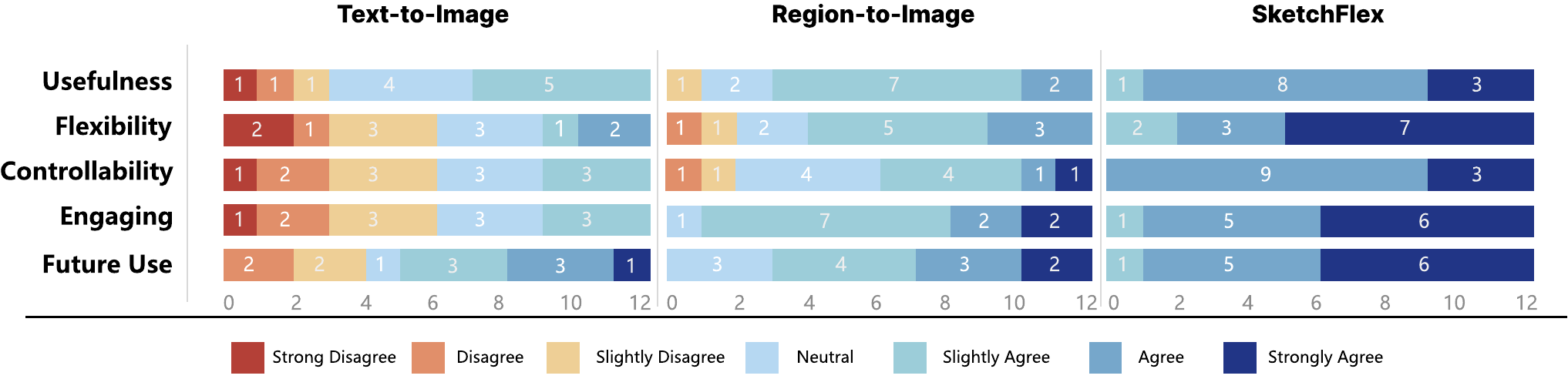}
    \vspace{-2mm}
    \caption{Overall usability ratings of system compared with two baselines.}
    \Description{Overall usability ratings of system compared with two baselines.}
    \label{fig:overall}
    \vspace{-3mm}
\end{figure*}

\textbf{Single Object Adjustment.} Single Object Adjustment during refinement provides flexible, fine-grained control, enabling iterative adjustments to better align the overall composition with user expectations, as shown in column 5. 
For example, in the first row, while maintaining the cat's fixed shape, its size was enlarged to better align with reference image 1 in Figure~\ref{fig:outcome_task1}.
In the second row, the train's size and location were slightly adjusted to more closely match reference image 2.
In the third row, the left person was moved downward slightly to increase the height gap between the two individuals, creating the impression that the right girl is jumping to defend the boy. 
We found that when the generated object did not perfectly match user expectations, all participants preferred using single object adjustment over adjusting their sketch. 
It indicates that single object adjustment was deemed a more flexible and stable way to achieve fine-grained refinements.

In most cases, users do not employ single object decomposition and adjustment as separate features but instead use them as complementary tools in an iterative generation process, as seen in column 6. 
For instance, in row 2, the user generated a new train and adjusted its position to stretch across the image. 
In row 3, the right girl was regenerated into a new shape and moved slightly to the right of the original mask position. 
Users alternated between modifying object shapes and adjusting their location or size until they were satisfied with the final result.

\subsubsection{System Overall Rating}\label{ssec:overallrating}

Figure~\ref{fig:overall} illustrates the user rating of the overall system comparison.

\textbf{Usefulness}. Most participants agreed that our system (mean = 6.16, SD = 0.55) was significantly more useful than both text-to-image (mean = 3.91, SD = 1.25) and region-based image generation methods (mean = 4.83, SD = 0.79).

\textbf{Flexibility}. In terms of flexibility, participants found that our system (mean = 6.41, SD = 1.60) offered far more input flexibility compared to text-to-image (mean = 3.50, SD = 1.60) and region-to-image (mean = 4.66, SD = 1.78). This flexibility stems from the ability to adjust not only text and sketches, but also individual objects, providing significantly more control over the image generation process (10/12 participants). As P5 noted, "\emph{With both text-to-image and region-based methods, there's always a part of the image that looks right and another part that doesn’t. But when I change the prompt or region, the whole image changes, making it difficult to get everything just right. This system solves that problem}."

\textbf{Controllability}. All participants agreed that \tool (mean = 6.25, SD = 0.43) provided better control than text-to-image (mean = 3.41, SD = 1.25) and region-based generation (mean = 4.50, SD = 1.26). 
\tool allows control over text and regions and to choose and fix specific shapes, enabling more precise adjustments.

\textbf{Engaging}. Participants also found our system to be more engaging (mean = 6.41, SD = 0.64) compared to text-to-image (mean = 3.41, SD = 1.25) and region-based methods (mean = 5.41, SD = 0.86). P6 remarked, "\emph{The various interaction methods made the process more engaging, helping me stay focused while iterating and refining the artwork to better match the imagery in my mind.}."

\textbf{Future use}. Lastly, participants saw greater future potential in our system (mean = 6.41, SD = 0.64) compared to text-to-image (mean = 4.50, SD = 1.61) and region-based generation (mean = 5.33, SD = 1.03), particularly for non-experts who could quickly create complete images with desired compositions (5/12 participants). However, a few participants (2/12) felt that the methods could have equally significant future potential.
% there was no significant difference in future potential among the methods. 
As P3 observed, "\emph{I think all the methods have future potential, but for different scenarios. Some areas require specific control, while others benefit from abstract control as inspiration. Overall, this system provides more options, especially in terms of controllability and ease of use}."

\section{Discussion}\label{sec:discussion}

\subsection{From Interactive Prompting to Interactive Multi-modal Prompting}
The rapid advancements of large pre-trained generative models including large language models and text-to-image generation models, have inspired many HCI researchers to develop interactive tools to support users in crafting appropriate prompts.
% Studies on this topic in last two years' HCI conferences are predominantly focused on helping users refine single-modality textual prompts.
Many previous studies are focused on helping users refine single-modality textual prompts.
However, for many real-world applications concerning data beyond text modality, such as multi-modal AI and embodied intelligence, information from other modalities is essential in constructing sophisticated multi-modal prompts that fully convey users' instruction.
This demand inspires some researchers to develop multimodal prompting interactions to facilitate generation tasks ranging from visual modality image generation~\cite{wang2024promptcharm, promptpaint} to textual modality story generation~\cite{chung2022tale}.
% Some previous studies contributed relevant findings on this topic. 
Specifically, for the image generation task, recent studies have contributed some relevant findings on multi-modal prompting.
For example, PromptCharm~\cite{wang2024promptcharm} discovers the importance of multimodal feedback in refining initial text-based prompting in diffusion models.
However, the multi-modal interactions in PromptCharm are mainly focused on the feedback empowered the inpainting function, instead of supporting initial multimodal sketch-prompt control. 

\begin{figure*}[t]
    \centering
    \includegraphics[width=0.9\textwidth]{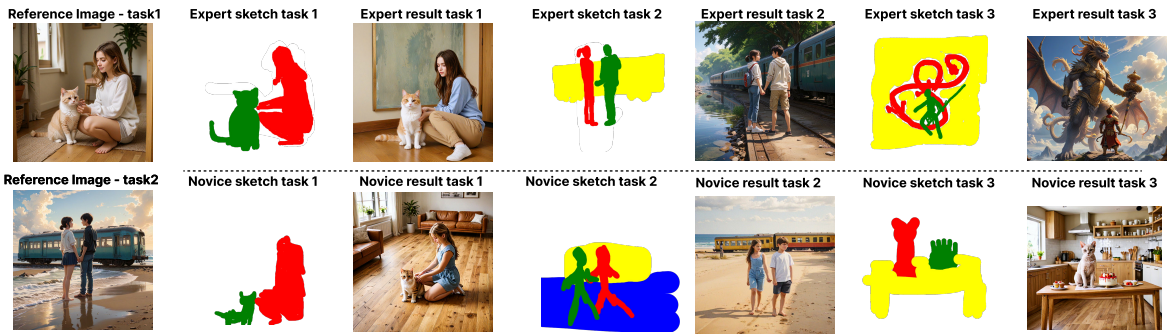}
    \vspace{-2mm}
    \caption{The comparison between novice and expert participants in painting reveals that experts produce more accurate and fine-grained sketches, resulting in closer alignment with reference images in close-ended tasks. Conversely, in open-ended tasks, expert fine-grained strokes fail to generate precise results due to \tool's lack of control at the thin stroke level.}
    \Description{The comparison between novice and expert participants in painting reveals that experts produce more accurate and fine-grained sketches, resulting in closer alignment with reference images in close-ended tasks. Novice users create rougher sketches with less accuracy in shape. Conversely, in open-ended tasks, expert fine-grained strokes fail to generate precise results due to \tool's lack of control at the thin stroke level, while novice users' broader strokes yield results more aligned with their sketches.}
    \label{fig:novice_expert}
    % \vspace{-3mm}
\end{figure*}

% In particular, in the initial control input, users are unable to explicitly specify multi-modal generation intents.
In another example, PromptPaint~\cite{promptpaint} stresses the importance of paint-medium-like interactions and introduces Prompt stencil functions that allow users to perform fine-grained controls with localized image generation. 
However, insufficient spatial control (\eg, PromptPaint only allows for single-object prompt stencil at a time) and unstable models can still leave some users feeling the uncertainty of AI and a varying degree of ownership of the generated artwork~\cite{promptpaint}.
% As a result, the gap between intuitive multi-modal or paint-medium-like control and the current prompting interface still exists, which requires further research on multi-modal prompting interactions.
From this perspective, our work seeks to further enhance multi-object spatial-semantic prompting control by users' natural sketching.
However, there are still some challenges to be resolved, such as consistent multi-object generation in multiple rounds to increase stability and improved understanding of user sketches.

\subsection{Novice Performance vs. Expert Performance}\label{sec:nVe}
In this section we discuss the performance difference between novice and expert regarding experience in painting and prompting.
First, regarding painting skills, some participants with experience (4/12) preferred to draw accurate and fine-grained shapes at the beginning. 
All novice users (5/12) draw rough and less accurate shapes, while some participants with basic painting skills (3/12) also favored sketching rough areas of objects, as exemplified in Figure~\ref{fig:novice_expert}.
The experienced participants using fine-grained strokes (4/12, none of whom were experienced in prompting) achieved higher IoU scores (0.557) in the close-ended task (0.535) when using \tool. 
This is because their sketches were closer in shape and location to the reference, making the single object decomposition result more accurate.
Also, experienced participants are better at arranging spatial location and size of objects than novice participants.
However, some experienced participants (3/12) have mentioned that the fine-grained stroke sometimes makes them frustrated.
As P1's comment for his result in open-ended task: "\emph{It seems it cannot understand thin strokes; even if the shape is accurate, it can only generate content roughly around the area, especially when there is overlapping.}" 
This suggests that while \tool\ provides rough control to produce reasonably fine results from less accurate sketches for novice users, it may disappoint experienced users seeking more precise control through finer strokes. 
As shown in the last column in Figure~\ref{fig:novice_expert}, the dragon hovering in the sky was wrongly turned into a standing large dragon by \tool.

Second, regarding prompting skills, 3 out of 12 participants had one or more years of experience in T2I prompting. These participants used more modifiers than others during both T2I and R2I tasks.
Their performance in the T2I (0.335) and R2I (0.469) tasks showed higher scores than the average T2I (0.314) and R2I (0.418), but there was no performance improvement with \tool\ between their results (0.508) and the overall average score (0.528). 
This indicates that \tool\ can assist novice users in prompting, enabling them to produce satisfactory images similar to those created by users with prompting expertise.

\subsection{Applicability of \tool}
The feedback from user study highlighted several potential applications for our system. 
Three participants (P2, P6, P8) mentioned its possible use in commercial advertising design, emphasizing the importance of controllability for such work. 
They noted that the system's flexibility allows designers to quickly experiment with different settings.
Some participants (N = 3) also mentioned its potential for digital asset creation, particularly for game asset design. 
P7, a game mod developer, found the system highly useful for mod development. 
He explained: "\emph{Mods often require a series of images with a consistent theme and specific spatial requirements. 
For example, in a sacrifice scene, how the objects are arranged is closely tied to the mod's background. It would be difficult for a developer without professional skills, but with this system, it is possible to quickly construct such images}."
A few participants expressed similar thoughts regarding its use in scene construction, such as in film production. 
An interesting suggestion came from participant P4, who proposed its application in crime scene description. 
She pointed out that witnesses are often not skilled artists, and typically describe crime scenes verbally while someone else illustrates their account. 
With this system, witnesses could more easily express what they saw themselves, potentially producing depictions closer to the real events. "\emph{Details like object locations and distances from buildings can be easily conveyed using the system}," she added.

\subsection{Progressive Sketching}
Currently \tool is mainly aimed at novice users who are only capable of creating very rough sketches by themselves.
However, more accomplished painters or even professional artists typically have a coarse-to-fine creative process. 
Such a process is most evident in painting styles like traditional oil painting or digital impasto painting, where artists first quickly lay down large color patches to outline the most primitive proportion and structure of visual elements.
After that, the artists will progressively add layers of finer color strokes to the canvas to gradually refine the painting to an exquisite piece of artwork.
One participant in our user study (P1) , as a professional painter, has mentioned a similar point "\emph{
I think it is useful for laying out the big picture, give some inspirations for the initial drawing stage}."
Therefore, rough sketch also plays a part in the professional artists' creation process, yet it is more challenging to integrate AI into this more complex coarse-to-fine procedure.
Particularly, artists would like to preserve some of their finer strokes in later progression, not just the shape of the initial sketch.
In addition, instead of requiring the tool to generate a finished piece of artwork, some artists may prefer a model that can generate another more accurate sketch based on the initial one, and leave the final coloring and refining to the artists themselves.
To accommodate these diverse progressive sketching requirements, a more advanced sketch-based AI-assisted creation tool should be developed that can seamlessly enable artist intervention at any stage of the sketch and maximally preserve their creative intents to the finest level. 

\subsection{Ethical Issues}
Intellectual property and unethical misuse are two potential ethical concerns of AI-assisted creative tools, particularly those targeting novice users.
In terms of intellectual property, \tool hands over to novice users more control, giving them a higher sense of ownership of the creation.
However, the question still remains: how much contribution from the user's part constitutes full authorship of the artwork?
As \tool still relies on backbone generative models which may be trained on uncopyrighted data largely responsible for turning the sketch into finished artwork, we should design some mechanisms to circumvent this risk.
For example, we can allow artists to upload backbone models trained on their own artworks to integrate with our sketch control.
Regarding unethical misuse, \tool makes fine-grained spatial control more accessible to novice users, who may maliciously generate inappropriate content such as more realistic deepfake with specific postures they want or other explicit content.
To address this issue, we plan to incorporate a more sophisticated filtering mechanism that can detect and screen unethical content with more complex spatial-semantic conditions. 
% In the future, we plan to enable artists to upload their own style model

% \subsection{From interactive prompting to interactive spatial prompting}

\subsection{Limitations and Future work}

    \textbf{User Study Design}. Our open-ended task assesses the usability of \tool's system features in general use cases. To further examine aspects such as creativity and controllability across different methods, the open-ended task could be improved by incorporating baselines to provide more insightful comparative analysis. 
    Besides, in close-ended tasks, while the fixing order of tool usage prevents prior knowledge leakage, it might introduce learning effects. In our study, we include practice sessions for the three systems before the formal task to mitigate these effects. In the future, utilizing parallel tests (\textit{e.g.} different content with the same difficulty) or adding a control group could further reduce the learning effects.

    \textbf{Failure Cases}. There are certain failure cases with \tool that can limit its usability. 
    Firstly, when there are three or more objects with similar semantics, objects may still be missing despite prompt recommendations. 
    Secondly, if an object's stroke is thin, \tool may incorrectly interpret it as a full area, as demonstrated in the expert results of the open-ended task in Figure~\ref{fig:novice_expert}. 
    Finally, sometimes inclusion relationships (\textit{e.g.} inside) between objects cannot be generated correctly, partially due to biases in the base model that lack training samples with such relationship. 

    \textbf{More support for single object adjustment}.
    Participants (N=4) suggested that additional control features should be introduced, beyond just adjusting size and location. They noted that when objects overlap, they cannot freely control which object appears on top or which should be covered, and overlapping areas are currently not allowed.
    They proposed adding features such as layer control and depth control within the single-object mask manipulation. Currently, the system assigns layers based on color order, but future versions should allow users to adjust the layer of each object freely, while considering weighted prompts for overlapping areas.

    \textbf{More customized generation ability}.
    Our current system is built around a single model $ColorfulXL-Lightning$, which limits its ability to fully support the diverse creative needs of users. Feedback from participants has indicated a strong desire for more flexibility in style and personalization, such as integrating fine-tuned models that cater to specific artistic styles or individual preferences. 
    This limitation restricts the ability to adapt to varied creative intents across different users and contexts.
    In future iterations, we plan to address this by embedding a model selection feature, allowing users to choose from a variety of pre-trained or custom fine-tuned models that better align with their stylistic preferences. 
    
    \textbf{Integrate other model functions}.
    Our current system is compatible with many existing tools, such as Promptist~\cite{hao2024optimizing} and Magic Prompt, allowing users to iteratively generate prompts for single objects. However, the integration of these functions is somewhat limited in scope, and users may benefit from a broader range of interactive options, especially for more complex generation tasks. Additionally, for multimodal large models, users can currently explore using affordable or open-source models like Qwen2-VL~\cite{qwen} and InternVL2-Llama3~\cite{llama}, which have demonstrated solid inference performance in our tests. While GPT-4o remains a leading choice, alternative models also offer competitive results.
    Moving forward, we aim to integrate more multimodal large models into the system, giving users the flexibility to choose the models that best fit their needs.

\section{Conclusion}\label{sec:conclusion}
In this paper, we present \tool, an interactive system designed to help novice users create high-quality, fine-grained images that align with their intentions based on rough sketches. 
The system first refines the user's initial prompt into a complete and coherent one that matches the rough sketch, ensuring the generated results are both stable, coherent and high quality.
To further support users in achieving fine-grained alignment between the generated image and their creative intent without requiring professional skills, we introduce a decompose-and-recompose strategy. 
This allows users to select desired, refined object shapes for individual decomposed objects and then recombine them, providing flexible mask manipulation for precise spatial control.
The framework operates through a coarse-to-fine process, enabling iterative and fine-grained control that is not possible with traditional end-to-end generation methods. 
Our user study demonstrates that \tool offers novice users enhanced flexibility in control and fine-grained alignment between their intentions and the generated images.

\begin{acks}
The authors wish to thank the anonymous reviewers for their valuable comments. 
This paper is partially supported by National Natural Science Foundation of China (NO. U23A20313, 62372471), the Guangzhou Basic and Applied Basic Research Foundation (No. 2024A04J6462), and The Science Foundation for Distinguished Young Scholars of Hunan Province (NO. 2023JJ10080).
\end{acks}

\balance
%%% -*-BibTeX-*-
%%% Do NOT edit. File created by BibTeX with style
%%% ACM-Reference-Format-Journals [18-Jan-2012].

\bibliographystyle{ACM-Reference-Format} 
\bibliography{chi25-714}
\appendix

\section{Appendix: Prompt}
\label{sec:appendix}
``Here is a sketch of an image. 
$\{input\_color\_mask\}$, while the rest of the white space is the background. 
I need you to infer details of the image based on the given sketch.
The details should include the possible background likely to be present with the $\{input\_color\_mask\}$, the attribute of each object (like wearing, texture, color etc.), the state (including action, posture, etc.) of each object, the direction of each object and the relationships between objects.

You should first analyze the mask carefully, considering the size, location, and relative position of each object mask. Ensure that specific actions are analyzed based on the mask, and infer each aspect with a reasoning process before providing the final output.
The final output format should be: $\{format\_example\}$, and you should refer to the example: $\{few\_shot\}$. You are going to complete the "" in each item, you need to complete them in multiple short phrases based on your above reasoning.

The state and relationship should be as detailed as possible while ensuring they align with the mask, formatted as: objectA action/spatial relation objectB, with both objectA and objectB included.
You should properly refer to some examples of attributes of object $\{attributes\}$ and relationships $\{relationships\}$.
Do not include words like `or', `possibly' in your final output, there should no ambiguity in your output.
Make sure all aspects of given mask is filled.''

\end{document}